\documentclass[10pt,a4paper,twocolumn]{article}

\usepackage[a4paper,left=1.8cm,right=1.8cm,top=2.2cm,bottom=2.4cm,columnsep=0.7cm]{geometry}

\usepackage{graphicx}%
\usepackage{multirow}%
\usepackage{amsmath,amssymb,amsfonts}%
\usepackage{amsthm}%
\usepackage{mathrsfs}%
\usepackage[title]{appendix}%
\usepackage[dvipsnames,table]{xcolor}%
\usepackage{textcomp}%
\usepackage{manyfoot}%
\usepackage{booktabs}%
\usepackage{algorithm}%
\usepackage{mhchem} 
\usepackage{algorithmicx}%
\usepackage{algpseudocode}%
\usepackage{listings}%
\usepackage{tikz}
\usepackage{comment}
\usepackage{subcaption}
\usepackage{caption}
\usepackage{threeparttable}
\graphicspath{{./Figures/}}

\definecolor{DarkBlackGreen}{RGB}{30,80,00}

\usepackage{fancyhdr}
\usepackage{titlesec}
\usepackage{cuted}
\usepackage[font=small,labelfont=bf,labelsep=period,justification=justified]{caption}
\usepackage{authblk}
\usepackage{cite}
\usepackage[colorlinks=true,linkcolor=RoyalBlue,citecolor=RoyalBlue,urlcolor=RoyalBlue,breaklinks=true]{hyperref}
\usepackage[switch]{lineno}

\theoremstyle{plain}%
\theoremstyle{definition}%

\titleformat{\section}{\normalfont\large\bfseries}{\thesection.}{0.5em}{}
\titleformat{\subsection}{\normalfont\normalsize\bfseries}{\thesubsection.}{0.5em}{}
\titleformat{\subsubsection}{\normalfont\normalsize\itshape}{\thesubsubsection.}{0.5em}{}
\titlespacing*{\section}{0pt}{1.3em plus 0.3em minus 0.2em}{0.7em}
\titlespacing*{\subsection}{0pt}{1.0em plus 0.2em minus 0.2em}{0.5em}
\titlespacing*{\subsubsection}{0pt}{0.8em plus 0.2em minus 0.2em}{0.4em}

\fancypagestyle{firstpage}{%
  \fancyhf{}%
  \fancyfoot[C]{\small\thepage}%
}
\usepackage{cleveref}

\crefname{figure}{Fig.}{Figs.}
\Crefname{figure}{Fig.}{Figs.}
\crefname{section}{Sec.}{Secs.}
\Crefname{section}{Sec.}{Secs.}
\crefname{subsection}{Sec.}{Secs.}
\Crefname{subsection}{Sec.}{Secs.}
\crefname{equation}{Eq.}{Eqs.}
\Crefname{equation}{Eq.}{Eqs.}

\crefname{table}{Tab.}{Tabs.}
\Crefname{table}{Tab.}{Tabs.}
\newcommand{\Vg}{V_\mathrm{g}}

\newcommand{\Rth}{R_{\mathrm{th}}}
\newcommand{\Cth}{C_{\mathrm{th}}}
\newcommand{\Tamb}{T_{\mathrm{amb}}}
\newcommand{\Ea}{E_{\mathrm{a}}}
\newcommand{\kb}{k_{\mathrm{B}}}

\newcommand{\dd}{\mathrm{d}}

\DeclareMathOperator{\sgn}{sgn}

\begin{document}

\twocolumn[
\begin{center}
{\LARGE\bfseries Current-Gated Nonlinear Dynamics of a Self-Heating Memristor: an Electrothermal Extension of the Pickett Filamentary Model\par}
\vspace{0.9em}
{\large 
N.G.~Koudafok\^e$^{1,2,\ast}$,\;
Florian~G\"unther$^{3}$,\;
Hilda~A.~Cerdeira$^{1}$,\;
A.~V.~Monwanou$^{2}$
\par}

\vspace{0.6em}

{\small
$^{1}$ICTP South American Institute for Fundamental Research,
Instituto de F\'{i}sica Te\'{o}rica (IFT--UNESP), Bloco~II,
Rua Dr.\ Bento Teobaldo Ferraz 271, Barra Funda,
S\~ao Paulo, 01140-070, Brazil\\

$^{2}$Institut de Math\'{e}matiques et de Sciences Physiques (IMSP),
Universit\'e d'Abomey Calavi (UAC),
Porto-Novo, B\'enin\\

$^{3}$S\~ao Paulo State University (UNESP),
Instituto de Geoci\^encias e Ci\^encias Exatas (IGCE),
Campus Rio Claro, Rio Claro -- SP, 13506--700, Brazil\\

$^{\ast}$Corresponding author:
\texttt{gilles.koudafoke@ictp-saifr.org}
\par}
\end{center}
\vspace{0.8em}
{\rule{\textwidth}{0.6pt}}
\vspace{0.4em}

\noindent{\bfseries Abstract.\ }%

Self-heating couples the electrical and thermal states of filamentary memristors. However, the widely used Pickett compact model of \ce{TiO2} resistive switching is isothermal and therefore cannot capture the resulting electrothermal dynamics. We introduce the Arrhenius-Thermal Filamentary Model (ATFM), which extends Pickett's tunneling-gap kinetics
by incorporating a dynamic heat balance and an Arrhenius-activated
switching rate. The resulting electrothermal feedback produces a sharp current-gated transition: below a critical drive current, the tunneling gap undergoes a non-returning ratchet drift, whereas above it, exponential locking of the filament kinetics establishes a bounded, drive-locked electrothermal oscillation. Using a stroboscopic Poincar\'e map and the Floquet multipliers of the resulting period-$1$ orbit, we characterize this onset as a threshold-like orbit contraction rather than
a classical local bifurcation. In the limit $\Ea\to0$, ATFM recovers the isothermal Pickett dynamics to numerical precision, as verified against an independent reference implementation over amplitude, frequency, and activation-energy sweeps. A variance-based Sobol' analysis with bootstrap confidence intervals identifies the excitation amplitude as the dominant
control parameter and the thermal resistance $\Rth$, rather than the
thermal capacitance $\Cth$, as the leading thermal contributor. A
geometry-dependent temperature constraint further reveals a non-monotonic operating window in which an intermediate active area maximizes the switching excursion. The predicted trajectories are reproduced by both a fully behavioral \texttt{SPICE} netlist and a \texttt{Verilog-A/OSDI} device implementation, making ATFM directly suitable for circuit simulation. Overall, ATFM reveals and realizes a self-heating-driven
dynamical regime within the widely used Pickett filamentary framework.

\vspace{0.5em}
\ce{TiO2} memristor compact model; Thermally activated
kinetics; Global sensitivity analysis (Sobol' indices).

\vspace{0.4em}
{\rule{\textwidth}{0.6pt}}
\vspace{1.2em}
]
\thispagestyle{firstpage}

\section{Introduction}

Since the theoretical prediction of the memristor by Chua in 1971 \cite{Chua1971}, its generalization to memristive systems \cite{Chua1976} and the experimental realization of a titanium-dioxide device by Strukov \textit{et al.} in 2008 \cite{Strukov2008}, memristors have attracted considerable interest for non-volatile memory, neuromorphic computing, and reconfigurable circuits \cite{Yang2013,Waser2007,Pershin2011,DiVentra2009}. In titanium dioxide, resistive switching is often attributed to the field-induced redistribution of oxygen vacancies, which alters the geometry of a conducting filament inside the oxide \cite{Yang2008,Schaub2003,Jeong2009}. In this process, a single normalized state variable is used to describe the drift-based model of Strukov \textit{et al.} \cite{Strukov2008}, while Pickett \textit{et al.}~\cite{Pickett2009,PickettThesis} proposed a more geometrical description based on the evolution of the nanometer-scale insulating gap between the conductive filament and the electrode. In this formulation, quantum-mechanical tunneling described by Simmons' theory \cite{Simmons1963a,Simmons1963b} is used to explicitly account for electron transport across the gap. Since then, the Pickett model has been widely used as a filamentary compact model of resistive switching of titanium-dioxide and has been adopted into \texttt{SPICE}-based frameworks \cite{Abdalla2011,Kolka2015}. Pickett’s model is essentially isothermal in the sense that temperature does not enter as a dynamical variable, and the switching-rate coefficients are treated as fixed material parameters, despite the geometrical description of the switching process. This is a major drawback since oxygen-vacancy migration is thermally activated with reported activation energies of about $0.2$--$0.8\,\mathrm{eV}$ \cite{DeLile2023,Li2012a,Janotti2010,Morgan2010,Morgan2009} and the tunneling current induces Joule heating in the active region. Therefore, thermal effects may alter the switching dynamics and device operation \cite{Ielmini2011,Valov2011,Kim2014,Mickel2014}.
The key missing ingredient is thus a self-consistent coupling between the filamentary gap dynamics, Joule self-heating, and temperature-dependent switching kinetics.

We present ATFM, an electrothermal extension of Pickett’s compact model, that introduces this coupling while staying within the original model structure. A dynamic thermal balance and an Arrhenius activation factor are added, but the original tunneling-current formulation, series-resistance regularization, and switching-kinetics structure are kept. The construction also retains Pickett’s model as a well-defined limit: in the limit of vanishing activation energy, ATFM recovers the temperature-independent Pickett dynamics. This limit is verified numerically by direct regression against an independently implemented reference model over independent sweeps of excitation amplitude, frequency, and activation energy. To validate circuit-level portability, the full model is also transcribed into a fully behavioral \texttt{SPICE} netlist, whose simulated trajectories reproduce the reference implementation at the baseline and extended operating points (Appendix~\ref{app:SPICE-validation}).

The resulting electrothermal framework exhibits a sharp current-gated transition between a non-returning ratchet drift and a bounded, drive-locked electrothermal oscillation. We show that this transition originates from the exponential locking structure of the filament kinetics and produces a localized enhancement of the response to ambient temperature. We further characterize this onset dynamically: a stroboscopic Poincar\'e map and the Floquet multipliers of the emerging period-$1$ orbit identify it as a threshold-like orbit contraction rather than a classical local bifurcation.

Using a variance-based global sensitivity (Sobol') analysis with bootstrap confidence intervals, we quantify the relative roles and interactions of the excitation amplitude, ambient temperature, activation energy, and thermal parameters. The excitation amplitude is the dominant control parameter, the activation energy acts mainly through its interaction with the drive near the ratchet-to-oscillation threshold, and the thermal resistance $\Rth$, rather than the thermal capacitance $\Cth$, is the leading thermal contributor to self-heating and hysteresis; peak Joule power and peak temperature, moreover, depend on excitation amplitude and ambient temperature through structurally different paths. Finally, the tangential extrapolation suggested by Kolka \textit{et al.} \cite{Kolka2015Enhanced} resolves the high-voltage ambiguity of the original Pickett port relation, recovering a monotonic tunneling characteristic without extending the practical operating range, which is instead bounded by a geometry-dependent overheating (thermal validity) limit. Analyzing this window across the active-area scale reveals a non-monotonic trade-off between thermal relaxation and kinetic accessibility, in which an intermediate active area maximizes the switching excursion within a prescribed temperature bound. Overall, ATFM reveals and realizes a self-heating-driven dynamical regime within the widely used Pickett filamentary framework.

The rest of this paper is organized as follows. The physical assumptions behind ATFM are described in \Cref{sec:physics}, the mathematical formulation in \Cref{sec:tr3b} and the physical parameters and their sources are summarized in \Cref{sec:parameters}. The mathematical and numerical properties of the model are discussed in \Cref{sec:properties}. The numerical validation, threshold analysis, and single-parameter sensitivity results are shown in \Cref{sec:numresults} and the variance-based global sensitivity analysis is reported in \Cref{sec:sobol-3b}. \Cref{sec:comparison} compares ATFM with the classical Pickett model, and \Cref{sec:limitations} discusses the limitations of the framework. Finally, \Cref{sec:conclusion} concludes with the main findings.

\section{Physical Structure and Electrothermal Modeling}
\label{sec:physics}

\begin{figure}[!ht!]
\centering \includegraphics[width=\columnwidth,height=4.5cm,trim=0 1cm 0 0cm,clip]{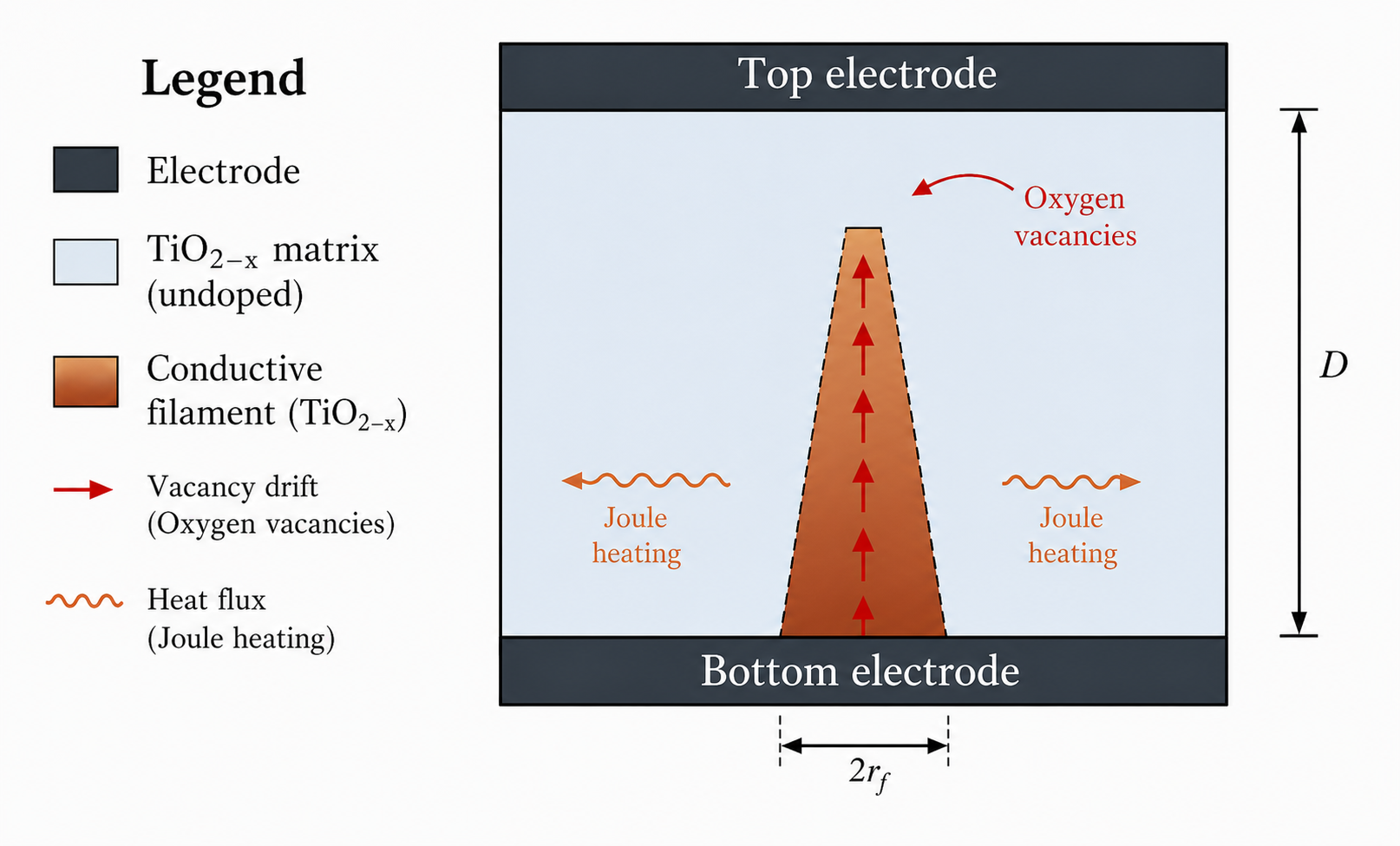}
  \vspace{-0.2cm}
  \caption{Physical structure of the titanium dioxide memristor considered
in ATFM. The conductive filament (gradient shading representing the
oxygen-vacancy concentration profile) connects both electrodes; Joule
heating generated inside the filament is dissipated through the
surrounding \ce{TiO2-x} matrix and the metallic electrodes.}
\label{fig:device}
\end{figure}

The considered device is a metal--insulator--metal (MIM) structure in
which a thin layer of oxygen-deficient titanium dioxide
(\ce{TiO2-x}) is sandwiched between two metallic electrodes
(\Cref{fig:device}). Under external electrical excitation, oxygen
vacancies drift within the oxide and progressively modify the geometry
of the conductive filament connecting the electrodes. During the SET
process, the filament grows toward the opposite electrode, reducing
the overall resistance, whereas RESET corresponds to a partial
rupture of the filament and an increase in resistance. This mechanism
has been extensively reported for titanium dioxide memristors and
provides the physical basis of the present model
\cite{Strukov2008,Pickett2009,Yang2013,Waser2007}.

The active region is described by a lumped electrothermal model with an
equivalent thermal resistance $\Rth$ and thermal capacitance $\Cth$, in
which the distributed heat equation is replaced by a single spatially
uniform temperature $T(t)$. Joule heating generated by the tunneling
current is assumed to be dissipated through the surrounding
\ce{TiO2-x} matrix and the metallic electrodes. This lumped
approximation is commonly used in compact electrothermal models to
capture the dominant thermal feedback while maintaining computational
efficiency \cite{Cahill2014}.

\begin{figure}[!ht!]
\centering
  \includegraphics[width=\columnwidth,height=3.5cm]{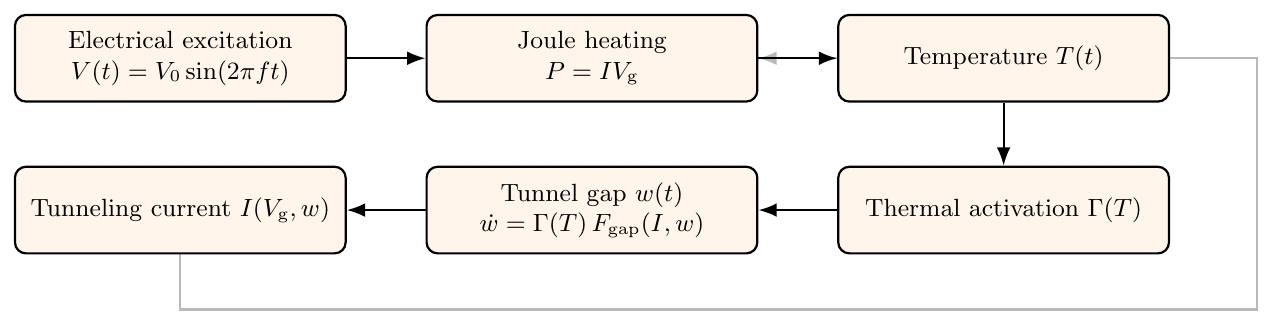}
  \vspace{-0.2cm}
  \caption{Electrothermal feedback loop of ATFM. The temperature-dependent
kinetics $\Gamma(T)$ couple the tunneling gap $w(t)$, tunneling current,
and Joule heating. The gap voltage $\Vg$ is obtained from
\Cref{eq:series} together with the tunneling relation.}
\label{fig:feedback}
\end{figure}

Within this framework, the temperature dependence of the filament
kinetics is introduced through an Arrhenius activation factor,
motivated by density-functional-theory (DFT) calculations of thermally activated oxygen-vacancy migration \cite{Janotti2010,Morgan2009,Morgan2010,DeLile2023,Li2012a}. An increase in the
applied voltage increases the tunneling current and hence the Joule
power, raising the device temperature. The resulting temperature
increase accelerates the switching kinetics through $\Gamma(T)$,
modifying the tunneling-gap width $w(t)$ and, consequently, the
tunneling current. Electrical transport, thermal balance, and gap
dynamics are therefore coupled through a nonlinear electrothermal
feedback loop (\Cref{fig:feedback}).

The model assumes that oxygen-vacancy migration, represented here through the evolution of the physical tunneling gap, is the dominant switching mechanism and that the associated kinetics follow an Arrhenius law. Electrical and thermal dynamics are fully coupled through the tunneling current, Joule heating, and temperature-dependent switching rate. Mechanical stresses, stochastic fluctuations, and structural degradation are neglected, while radiative and thermoelectric contributions to heat transfer are assumed negligible compared with
Joule heating. All material parameters are taken as constant unless an explicit temperature dependence is introduced in the model. These assumptions define the scope of ATFM and provide the physical basis for the mathematical formulation presented in \Cref{sec:tr3b}.


\section{Mathematical Formulation of ATFM}
\label{sec:tr3b}

The normalized state variable used by drift-based compact models such as Strukov's original formulation provides a phenomenological description
of the conductive filament but does not represent its microscopic
geometry. ATFM instead builds directly on the physically based
compact model of Pickett \textit{et al.} \cite{Pickett2009,PickettThesis},
in which switching is represented by the evolution of the nanoscale insulating gap separating the conductive filament from the metallic electrode, while electron transport is described by quantum tunneling following Simmons' generalized theory
\cite{Simmons1963a,Simmons1963b}.

\subsection{State Variable and Electrical Transport}

The internal state variable is the instantaneous tunneling-gap width $w(t)$, expressed in nanometers. Small values of $w$ correspond to a nearly continuous filament and the low-resistance state (LRS), whereas larger values correspond to filament rupture and the high-resistance
state (HRS).

Following Pickett's parameterization of the Simmons tunneling formula
\cite{Pickett2009,PickettThesis}, the effective tunneling barrier is characterized by the auxiliary width,

\begin{equation}
w_2 = w_1 + w - \frac{0.9183}{2.85+4\lambda-2|\Vg|},
\qquad
\lambda=\frac{L_m}{w},
\label{eq:w2}
\end{equation}
where $\Vg$ is the voltage across the gap, and $w_1$ and $L_m$ are
fitted constants \cite{PickettThesis}. Defining
$\Delta w=w_2-w_1$, the mean tunneling barrier height is
\begin{equation}
\begin{split}
    \phi_I = \phi_0
-|\Vg|\,\frac{w_1+w_2}{2w}-\\
1.15\,\lambda\,w\,
\frac{\ln\!\big[(w_2/w_1)(w-w_1)/(w-w_2)\big]}{\Delta w},
\end{split}
\label{eq:phiI}
\end{equation}

with $\phi_0$ the zero-field barrier height. The tunneling current is
then given by
\begin{equation}
\begin{split}
I =\sgn(\Vg)\,\frac{0.0617}{\Delta w^2}&
\Big[
\phi_I e^{-B\sqrt{\phi_I}}
-(\phi_I+|\Vg|)\,
e^{-B\sqrt{\phi_I+|\Vg|}}
\Big],\\
\qquad
B=10.246\,\Delta  w&,    
\end{split}
\label{eq:simmons}
\end{equation}

\paragraph{Correction of the large-voltage ambiguity.}
The original Pickett port relation exhibits a known ambiguity at large $|\Vg|$, where the denominator in \Cref{eq:w2} can approach zero and produce non-physical negative resistance and numerical non-convergence. Kolka, Biolek, and Biolkov\'a attributed this behavior to the unconditional use of a low-voltage approximation for $w_2$ outside the range in which Simmons' derivation applies \cite{Kolka2015}.

They proposed both an exact solution of the underlying cubic using trigonometric substitution \cite{Kolka2015} and a simpler tangential extrapolation of the static $I$--$\Vg$ characteristic, which is adopted here \cite{Kolka2015Enhanced}. Below the threshold
\begin{equation}
\begin{split}
\Vg^0(w)=\alpha +\beta\,(w-1.228) 
\quad[\mathrm{V},\ w\ \mathrm{in\ nm}],
\label{eq:vg0kolka}\\
\text{with $\alpha = 0.9$ and $\beta=0.36$} 
\end{split}
\end{equation}
\Cref{eq:w2,eq:phiI,eq:simmons} are used unchanged. Above this
threshold, the current is extrapolated exponentially, with continuity
of both value and logarithmic slope:
\begin{equation}
\begin{split}
I(\Vg)
=&
I\big(\sgn(\Vg)\Vg^0\big)
\exp\!\Big[k\big(|\Vg|-\Vg^0\big)\Big]\sgn(\Vg),\\
\qquad
k(w)=&
\left.
\frac{d\ln|I|}{d|\Vg|}
\right|_{|\Vg|=\Vg^0(w)},
\end{split}
\label{eq:kolka-extrap}
\end{equation}

where $k(w)$ is evaluated numerically from \Cref{eq:simmons} at
$|\Vg|=\Vg^0(w)$. This construction preserves the original characteristic below $\Vg^0(w)$ and provides a monotone continuation above it, removing the ambiguity in solving the port equation. The resulting characteristic remains smooth and monotone over the tested range up to $|\Vg|=2$~V, whereas the uncorrected relation diverges or becomes non-physical. As discussed in \Cref{sec:numresults,sec:limitations},
this numerical treatment does not by itself extend the useful
operating range, which remains constrained by a geometry-dependent temperature-bound validity limit.

The applied voltage is shared between the tunneling gap and the series resistance of the electrodes and filament,
\begin{equation}
V=\Vg+R_sI,
\label{eq:series}
\end{equation}
which is solved simultaneously with the tunneling relation to obtain $\Vg$ for a given applied voltage. The series resistance provides a negative electrical feedback: increasing current increases the voltage drop across $R_s$, thereby reducing $\Vg$ and limiting the current. This regularizing mechanism is retained explicitly in ATFM
(\Cref{sec:numresults}).

\subsection{Electrothermal Gap Dynamics}

The original Pickett model assumes temperature-independent
switching-rate coefficients $f_{\rm on}$ and $f_{\rm off}$. Here, these coefficients are multiplied by an Arrhenius factor motivated by the thermally activated nature of oxygen-vacancy migration \cite{Janotti2010,Morgan2010,DeLile2023,Li2012a}:
\begin{equation}
\Gamma(T)=
\exp\!\left[
-\frac{\Ea}{\kb}
\left(
\frac{1}{T}-\frac{1}{T_0}
\right)
\right],
\label{eq:thermalfactor}
\end{equation}
which gives the thermally activated gap dynamics.
\begin{equation}
\frac{dw}{dt}
=
\Gamma(T)\,F_{\rm gap}(I,w).
\label{eq:gapthermal}
\end{equation}
Here, $F_{\rm gap}$ retains the hyperbolic-sine/exponential structure of the original Pickett kinetics, with distinct expressions for $I>0$ and $I<0$ (the full expression, which follows Pickett's original formulation
\cite{Pickett2009,PickettThesis}, is given in \Cref{Eq_Complete}; the parameters $i_{\rm on}$, $i_{\rm off}$, $a_{\rm on}$, $a_{\rm off}$, and $w_c$ are listed in \Cref{sec:parameters}).

The temperature evolves according to the lumped thermal model
introduced in \Cref{sec:physics}, with the dissipated power associated
with the tunneling gap:
\begin{equation}
\Cth\frac{dT}{dt}
=
I\Vg-\frac{T-\Tamb}{\Rth}.
\label{eq:thermalgap}
\end{equation}
Thus, the gap dynamics and thermal balance are coupled through the tunneling current and gap voltage: the electrical state determines the Joule power, while the resulting temperature modifies the switching
kinetics through $\Gamma(T)$.

\subsection{Complete ATFM System}

The complete electrothermal model is therefore defined by
\begin{equation}
\left\{
\begin{aligned}
I &= f(\Vg,w)
&& \text{(\Cref{eq:w2,eq:phiI,eq:simmons})},\\
V &= \Vg+R_sI,\\
\frac{dw}{dt} &= \Gamma(T)\,F_{\rm gap}(I,w),\\
\Cth\frac{dT}{dt}
&= I\Vg-\frac{T-\Tamb}{\Rth}.
\end{aligned}
\right.
\label{eq:system3b}
\end{equation}

ATFM retains Pickett's tunneling-current formulation and
series-resistance regularization while introducing a dynamic coupling between the gap kinetics and temperature through the Arrhenius factor and the Joule-heating balance. This coupling results in an implicitly constrained electrothermal system: the tunneling relation must be solved together with the series-resistance equation to determine $\Vg$ for each state of the gap and thermal dynamics. The mathematical and numerical properties of this coupled system are examined in \Cref{sec:properties}, followed by the validation, threshold analysis,
and sensitivity studies presented in \Cref{sec:numresults}.


\section{Physical Parameters and Model Calibration}
\label{sec:parameters}

All model parameters have been selected according to experimentally validated literature whenever possible. Parameters introduced specifically for the proposed electrothermal formulation are explicitly identified and physically justified. Values are grouped into geometrical parameters (\Cref{tab:geometry}), filamentary parameters inherited from Pickett's original calibration (\Cref{tab:pickett}), thermal parameters (\Cref{tab:thermal}), and parameters introduced by the present work (\Cref{tab:new}).

\begin{table*}[!t]
\centering
\caption{Geometrical parameters.}
\label{tab:geometry}
\begin{tabular}{lllll}
\hline
Parameter & Symbol & Value & Unit & Origin\\
\hline
Oxide thickness & $D$ & $10$ & nm & Strukov \textit{et al.} \cite{Strukov2008}\\
Reference active section & $A_b$ & $50\times50$ & nm$^2$ & Fabricated junction size \cite{PickettThesis,Yang2008,Miao2011}$^{*}$\\
\hline
\end{tabular}
\begin{flushleft}
\footnotesize $^{*}$The switching-kinetics parameters used throughout this work are Pickett's own \cite{PickettThesis}, extracted from a $5\times5\,\mu$m$^2$ cross-point device deliberately chosen to minimize the parasitic series resistance during time-sampled measurements.
Pickett reports that the dynamical switching behavior is relatively insensitive to device size-—because electroforming localizes conduction to a single filament--and reproduces the same analysis on devices down to $50\times50$~nm$^2$ defined by imprint lithography \cite{PickettThesis,Yang2008}. Since the kinetic parameters are filament-localized rather than junction-area-dependent, the reference thermal section is set to this smallest fabricated junction, $A_b=50\times50$~nm$^2$, which also coincides with the crossbar junction size reported for related HP Labs \ce{TiO2} devices
\cite{Miao2011}.
\end{flushleft}
\end{table*}

\begin{table*}[!t]
\centering
\caption{Filamentary parameters of ATFM, from Pickett's thesis \cite{PickettThesis} and consistently reproduced in independent \texttt{SPICE} implementations \cite{Abdalla2011,Kolka2015}.}
\label{tab:pickett}
\begin{tabular}{lllll}
\hline
Parameter & Symbol & Value & Unit & Origin\\
\hline
Barrier height & $\phi_0$ & $0.95$ & V & Pickett \cite{PickettThesis}\\
Characteristic length & $L_m$ & $0.0998$ & nm & Pickett \cite{PickettThesis}\\
Reference width & $w_1$ & $0.1261$ & nm & Pickett \cite{PickettThesis}\\
Critical gap & $w_c$ & $0.107$ & nm & Pickett \cite{PickettThesis}\\
ON current parameter & $i_{\rm on}$ & $8.9\pm0.3$ & $\mu$A & Pickett \cite{PickettThesis}\\
OFF current parameter & $i_{\rm off}$ & $115\pm4$ & $\mu$A & Pickett \cite{PickettThesis}\\
ON gap parameter & $a_{\rm on}$ & $1.8\pm0.01$ & nm & Pickett \cite{PickettThesis}\\
OFF gap parameter & $a_{\rm off}$ & $1.2\pm0.02$ & nm & Pickett \cite{PickettThesis}\\
ON switching rate & $f_{\rm on}$ & $40\pm10$ & $\mu$m/s & Pickett \cite{PickettThesis}\\
OFF switching rate & $f_{\rm off}$ & $3.5\pm1$ & $\mu$m/s & Pickett \cite{PickettThesis}\\
Switching-rate saturation & $b$ & $500\pm80$ & $\mu$A & Pickett \cite{PickettThesis}\\
Series resistance & $R_s$ & $215$ & $\Omega$ & Pickett \cite{PickettThesis}\\
Admissible gap range & $[w_{\min},w_{\max}]$ & $[1.0,2.0]$ & nm & Abdalla \& Pickett \cite{Abdalla2011}\\
\hline
\end{tabular}
\end{table*}

\begin{table*}[!t]
\centering
\caption{Thermal parameters used in ATFM. The thermal resistance
$\Rth$ and thermal capacitance $\Cth$ are \emph{derived} from the
material properties and the assumed active geometry
(\Cref{tab:geometry}), rather than measured directly on an individual
device. The molar heat capacity $C_{p,\rm m}$ is converted to a
mass-specific heat capacity through the molar mass
$M_{\mathrm{TiO_2}}$.}
\label{tab:thermal}
\begin{tabular}{lllll}
\hline
Parameter & Symbol & Value & Unit & Origin\\
\hline
Thermal conductivity (\ce{TiO2-x})
& $\kappa$
& $1.6$
& W\,m$^{-1}$K$^{-1}$
& Mun \textit{et al.} \cite{Mun2007TiO2ThermalCond}, Liu \textit{et al.} \cite{Liu2017TiO2x}\\

Mass density (rutile)
& $\rho$
& $4250$
& kg\,m$^{-3}$
& Standard value\\

Molar heat capacity
& $C_{p,\rm m}$
& $55$
& J\,mol$^{-1}$K$^{-1}$
& Smith \& Carpenter \cite{Smith2009}\\

Molar mass of \ce{TiO2}
& $M_{\mathrm{TiO_2}}$
& $79.866\times10^{-3}$
& kg\,mol$^{-1}$
& Standard molecular mass\\

Equivalent active area
& $A_b$
& $50\times50$
& nm$^2$
& Fabricated junction, \Cref{tab:geometry}\\

Oxide thickness
& $D$
& $10$
& nm
& \Cref{tab:geometry}\\

Thermal resistance
& $\Rth$
& $2.50\times10^{6}$
& K/W
& Derived ($D/\kappa A_b$)\\

Thermal capacitance
& $\Cth$
& $7.32\times10^{-17}$
& J/K
& Derived from $\dfrac{\rho C_{p,\rm m}}{M_{\mathrm{TiO_2}}}A_bD$\\

Reference temperature
& $T_0$
& $293$
& K
& Present work\\

Ambient temperature
& $\Tamb$
& $293$
& K
& Present work\\

Activation energy
& $\Ea$
& $0.7$
& eV
& Representative value \cite{Morgan2010,Janotti2010}\\

Boltzmann constant
& $\kb$
& $8.617\times10^{-5}$
& eV/K
& CODATA\\
\hline
\end{tabular}
\end{table*}

\begin{table*}[!t]
\centering
\caption{Parameters and modeling assumptions introduced in ATFM.}
\label{tab:new}
\begin{tabular}{llll}
\hline
Quantity & Purpose & Origin & Status\\
\hline
Dynamic temperature $T(t)$ & Electrothermal coupling & Present work & Introduced here (ATFM extension)\\
$\Gamma(T)$-activated gap kinetics & Thermal activation & Present work & Introduced here (ATFM extension)\\
Thermal RC network ($\Rth,\Cth$) & Transient self-heating & Present work & Modeling assumption\\
Equivalent active section $A_b$ & Thermal calibration & Present work & Fabricated junction size \cite{PickettThesis,Yang2008,Miao2011}\\
\hline
\end{tabular}
\end{table*}

The resulting thermal time constant, $\tau_{\rm th}=\Rth\Cth\approx1.8\times10^{-10}\,\mathrm{s}$, is several orders of magnitude smaller than any electrically driven excitation period considered in this work (\Cref{sec:properties}). The thermal resistance itself, $\Rth=D/(\kappa A_b)$, is a one-dimensional effective parameter that lumps the electrodes, the interface thermal resistance, lateral heat diffusion, the spreading resistance of the localized filamentary hot spot, and a thermal cross-section potentially distinct from the conduction section into a single geometry-dependent calibration constant rather than a first-principles value, consistent with its role as the principal thermal contributor identified in \Cref{sec:sobol-3b}. Absolute temperatures consequently scale with the chosen reference section. The current-gated switching regimes (ratchet versus bounded oscillation) are set by the Pickett kinetic current scale and are insensitive to this calibration choice; the thermal operating window itself, by contrast, depends on the active area, as quantified in \Cref{sec:kolka-window}. The activation energy $\Ea=0.7$~eV lies within, but does not pin down precisely, the DFT range $0.19$--$0.82$~eV reported for oxygen-vacancy migration in rutile \ce{TiO2} \cite{Morgan2010,Janotti2010,DeLile2023,Li2012a}; this parameter should be treated as representative rather than calibrated for a specific device.

\section{Mathematical and Numerical Properties}
\label{sec:properties}

ATFM defines a nonlinear electrothermal dynamical system with state vector $\mathbf X(t)=[w(t),T(t)]^{\!\top}$. For a given applied excitation, the tunneling voltage $\Vg$ is determined implicitly from the port relation \Cref{eq:series} together with the tunneling characteristic \Cref{eq:simmons}. Within the admissible operating domain, the Kolka-corrected characteristic is continuous and monotone, so that this algebraic relation admits a unique local solution for $\Vg$ as a function of the instantaneous state. The resulting right-hand side is continuously differentiable away from the
singular region discussed in \Cref{sec:tr3b}, since $\Gamma(T)$ is $C^1$ for $T>0$ and the thermal balance is continuously differentiable. The Picard--Lindel\"of theorem \cite{Khalil2002,Strogatz2024} therefore guarantees a unique local solution for prescribed initial conditions and excitation.

The gap is constrained to
$0<w_{\min}\leq w(t)\leq w_{\max}$ (\Cref{tab:pickett}). The positivity of the temperature, $T(t)>0$, is likewise preserved for any admissible trajectory. From \Cref{eq:simmons}, $I$ and $\Vg$ have the same sign, so that the dissipated power satisfies $I\Vg\geq0$. Consequently,
\Cref{eq:thermalgap} gives
\begin{equation}
\Cth\dot T\big|_{T=0}
=
I\Vg+\frac{\Tamb}{\Rth}
>0,
\label{eq:tpositivity}
\end{equation}
whenever $\Tamb>0$. The vector field therefore points into the region $\{T>0\}$ at the boundary $T=0$, so this region is forward invariant: $T(t)>0$ for all $t$ given $T(0)>0$, independently of the simultaneous evolution of $w$.

The characteristic thermal time-scale,
\begin{equation}
\tau_{\rm th}=\Rth\Cth
\approx1.8\times10^{-10}\,\mathrm{s},
\label{eq:tauth}
\end{equation}
(\Cref{tab:thermal}) is much shorter than the electrical excitation
periods considered here ($0.5$--$2$~Hz, corresponding to periods of
approximately $0.5$--$2$~s). This strong separation of time-scales
makes the coupled system numerically stiff and motivates the use of
an implicit integration scheme. Physically, it also implies that the
temperature responds rapidly to variations in the dissipated power,
rather than developing an independent slow thermal hysteresis. This
quasi-static thermal response is confirmed numerically in
\Cref{sec:numresults}: within the voltage range studied there,
transient temperature excursions synchronized with the voltage peaks
relax toward $\Tamb$ near each zero crossing rather than accumulating
from cycle to cycle. Because the temperature returns to $\Tamb$ each
cycle and does not grow from one cycle to the next, the periodic thermal
response remains bounded in the range studied. At larger drive
amplitudes and active areas the model instead reaches a
geometry-dependent overheating threshold---a temperature-bound validity
limit identified in \Cref{sec:numresults}---beyond which the
constant-property assumptions of the lumped thermal model cease to hold.

The system is therefore integrated using the implicit Runge--Kutta
Radau IIA method \cite{Hairer2015,Radau1985}, with adaptive time
stepping and tight relative and absolute tolerances. At each
right-hand-side evaluation, \Cref{eq:series} is solved for $\Vg$,
with $I=I(\Vg,w)$ given by \Cref{eq:simmons}, using a bracketed
root-finding method restricted to $|\Vg|<2.2$~V. With the
Kolka-corrected tunneling relation described in \Cref{sec:tr3b}, this
bound is no longer imposed by the original numerical singularity: the
corrected characteristic remains monotone and solvable throughout the
tested range. Instead, the upper bound marks the point beyond which
the exponential extrapolation itself is no longer regarded as a
physically meaningful representation of the device.

These mathematical and numerical properties determine the integration
strategy used throughout the simulations and provide the basis for the
validation, threshold, and sensitivity analyses presented in
\Cref{sec:numresults}.

\section{Numerical Validation and Results}
\label{sec:numresults}

ATFM was simulated under sinusoidal voltage excitation,
$V(t)=V_0\sin(2\pi ft)$, starting from thermal equilibrium
$T(0)=\Tamb$. Unless otherwise stated, the baseline operating point is
$V_0=0.8$~V, $f=1$~Hz, and $\Ea=0.7$~eV, with the excitation amplitude
maintained within the admissible voltage range discussed in
\Cref{sec:tr3b}.

\begin{figure}[!ht]
\centering
\includegraphics[width=7cm,height=4.5cm,trim=0 0 0 1.8cm, clip]{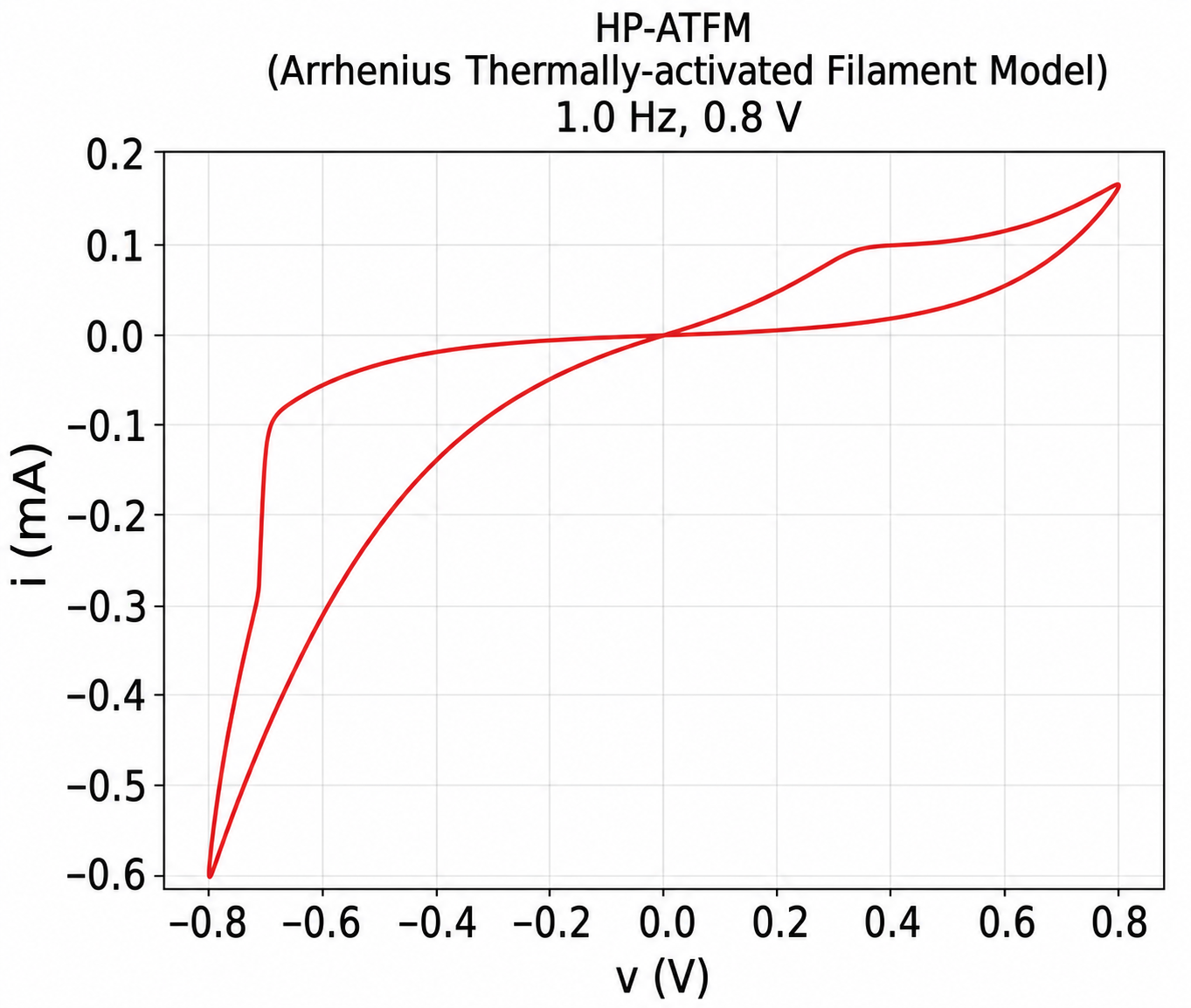}
\caption{Current--voltage hysteresis loop obtained with ATFM at the
baseline operating point ($V_0=0.8$~V, $f=1$~Hz, $\Ea=0.7$~eV), shown over the last simulated period (\texttt{SPICE} reproduction in
Appendix~\ref{app:SPICE-validation}).}
\label{fig:results}
\end{figure}

At this operating point, the current--voltage trajectory over the last
simulated period exhibits a bounded hysteretic response
(\Cref{fig:results}). The tunneling gap oscillates between approximately
$1.20$ and $1.50$~nm, while the active-region temperature develops
narrow transient peaks reaching approximately $1300.7$~K, synchronized
with the current maxima (\Cref{fig:w-two-points}). The temperature subsequently
relaxes toward $\Tamb$ near each zero crossing, consistent with the
quasi-static thermal regime established in \Cref{sec:properties}.

The predicted temperature scale is comparable to hot-spot temperatures
reported during resistive switching in the RRAM literature
\cite{Ielmini2011,Kim2014}. This comparison is qualitative; however,
because the present model uses a lumped thermal description, the
absolute temperature depends strongly on the assumed active area
$A_b$ (\Cref{tab:geometry}). The influence of this geometric parameter
is examined separately in \Cref{sec:kolka-window}.

\subsection{Validation of the Quasi-Static Thermal Approximation}
\label{sec:quasistatic-validation-3b}

The characteristic thermal timescale derived in
\Cref{sec:properties},
\begin{equation}
\tau_{\rm th}
=\Rth\Cth
\approx1.8\times10^{-10}\,\mathrm{s},
\end{equation}
is many orders of magnitude shorter than the electrical excitation
period. The leading-order thermal response can therefore be described
by the quasi-static relation
\begin{equation}
T(t)\simeq T_{\rm approx}(t)
\equiv
\Tamb+\Rth P_{\rm Joule}(t),
\label{eq:quasistatic}
\end{equation}
obtained by setting $\dot T\simeq0$ in \Cref{eq:thermalgap}.

The relevant small parameter is more precisely determined by the
timescale of the Joule-power evolution,
\begin{equation}
\varepsilon
\equiv
\frac{\tau_{\rm th}}{\tau_P},
\qquad
\tau_P
\equiv
\frac{P_{\max}}
{|\dot P_{\rm Joule}|_{\max}},
\label{eq:epsilon-3b}
\end{equation}
rather than directly by the forcing frequency. This formulation is
important because the nonlinear tunneling dynamics generate strongly
non-sinusoidal power transients even under sinusoidal voltage
excitation.

\begin{figure}[!ht]
\includegraphics[width=8.5cm,height=4.5cm,trim=0 0 0 0cm, clip]{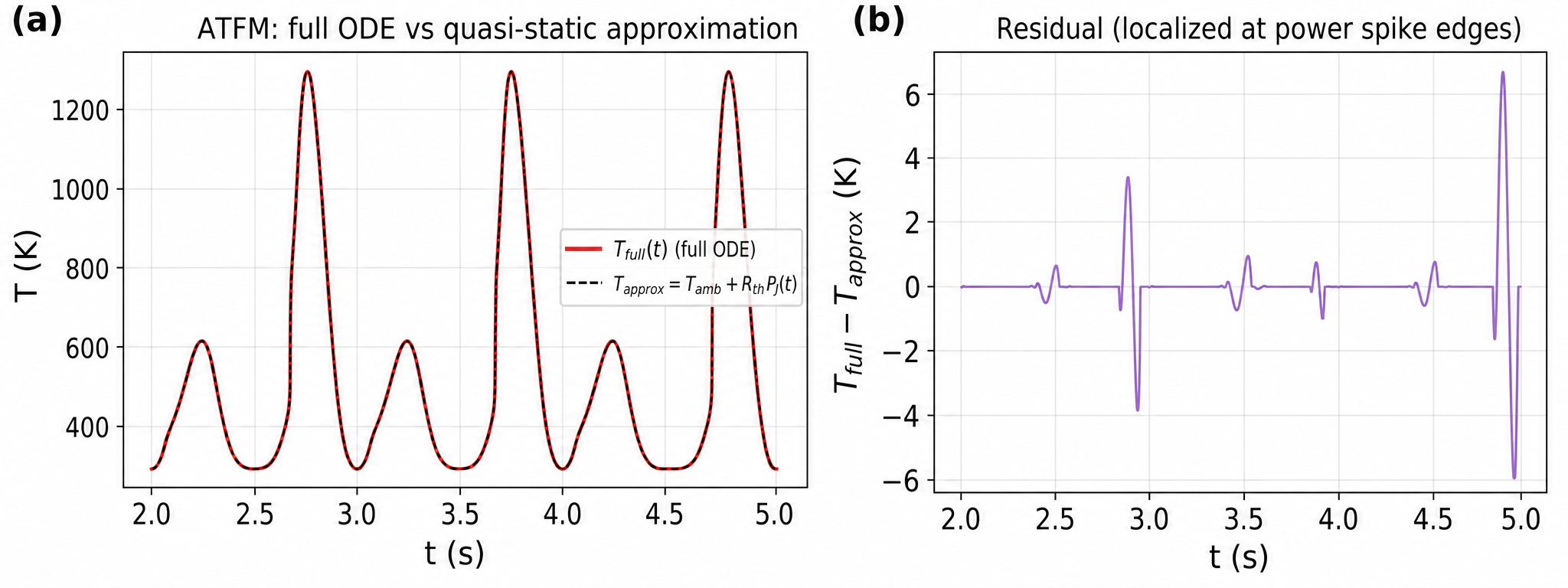}
\caption{ATFM: comparison between the full thermal solution
$T_{\rm full}(t)$ and the quasi-static approximation
$T_{\rm approx}(t)=\Tamb+\Rth P_{\rm Joule}(t)$ at
$V_0=0.8$~V, $f=1$~Hz, and $\Ea=0.7$~eV (a). The residual (b)
reaches $4.4$~K, corresponding to $0.44\,\%$ of the peak temperature rise, and is concentrated at the steepest edges of the Joule-power spikes. (\texttt{SPICE} comparison reproduced in Appendix~\ref{app:SPICE-validation}).}
\label{fig:quasistatic-3b}
\end{figure}

At the reference operating point,
\Cref{fig:quasistatic-3b} compares the full thermal solution
$T_{\rm full}(t)$ with the algebraic approximation evaluated from the
same simulated Joule-power trajectory. The two responses are nearly
indistinguishable over the full cycle. The maximum residual is
$4.4$~K, corresponding to only $0.44\,\%$ of the peak temperature
rise $\Delta T\approx1008$~K. The residual is localized at the
steepest portions of the power spikes and therefore represents a
small but finite thermal lag associated with the thermal capacitance.

For the reference operating point,
$\tau_P\approx0.018$~s and
$\varepsilon\approx10^{-8}$. The very small value of $\varepsilon$
confirms that the thermal response remains deep in the quasi-static regime despite the abrupt variation of $P_{\rm Joule}(t)$. The same diagnostic near the ratchet-to-oscillation threshold gives an even
smaller value, $\varepsilon\approx3\times10^{-9}$ at $V_0=0.70$~V, with similarly small residuals at $V_0=0.65$~V. Thus, proximity to the switching threshold does not by itself compromise the quasi-static approximation.

The comparison demonstrates that the algebraic reduction in
\Cref{eq:quasistatic} reproduces the full thermal dynamics to well below $1\,\%$ at the reference operating point. Nevertheless, the
dynamic thermal equation is retained throughout the simulations
because it remains valid when $\tau_{\rm th}$ is no longer negligible,
as examined explicitly in \Cref{sec:Cth-3b}.

\subsection{Operating-Point Sensitivity of ATFM}
\label{sec:hptr3b-sensitivity}

\begin{figure}[!ht]
\centering
\includegraphics[width=8cm,height=7cm,trim=0 0 0 0cm, clip]{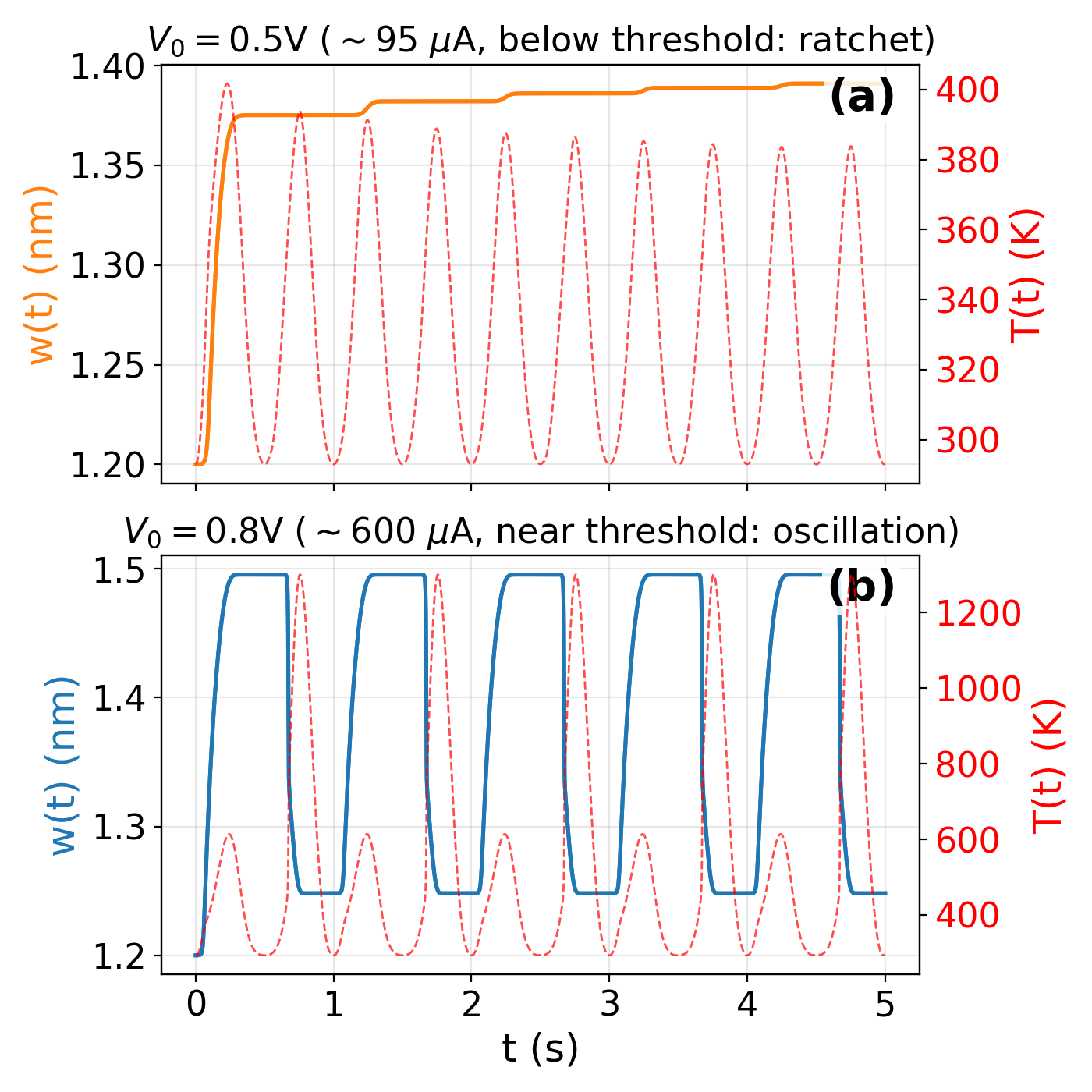}
\caption{ATFM: simultaneous evolution of $w(t)$ and $T(t)$ at
$V_0=0.5$~V ($I_0\approx95$~$\mu$A, $(a)$) and $V_0=0.8$~V
($I_0\approx600$~$\mu$A, $(b)$), $f=1$~Hz, and $\Ea=0.7$~eV.
The lower current produces a non-returning ratchet trajectory,
whereas the higher current produces a bounded oscillation.}
\label{fig:w-two-points}
\end{figure}

The bounded oscillatory behavior observed at the reference operating
point results from the nonlinear threshold structure of the filament
kinetics. \Cref{fig:w-two-points} shows the simultaneous evolution of the
tunneling gap and active-region temperature. 

The current amplitudes associated with the Pickett calibration,
$i_{\rm on}\sim9\,\mu$A and $i_{\rm off}\sim115\,\mu$A, strongly
influence the accessibility of the two switching branches. To
quantify this dependence, \Cref{fig:w-two-points} compares two voltage
amplitudes. At $V_0=0.5$~V, corresponding to
$I_0\approx95\,\mu$A, the gap increases in a staircase-like,
non-returning trajectory (\Cref{fig:w-two-points}(a)). At $V_0=0.8$~V,
$I_0\approx600\,\mu$A, the gap instead undergoes a bounded oscillation
between approximately $1.20$ and $1.50$~nm, while narrow temperature
spikes reach $1300.7$~K (\Cref{fig:w-two-points}(b)).

This qualitative change originates from the exponential lock
structure of $F_{\rm gap}(I,w)$ in \Cref{eq:gapthermal}. Each
switching branch contains a factor
\[
\exp\!\left[-\exp\!\left(\frac{\cdot}{w_c}\right)
-\frac{w}{w_c}\right],
\]
which suppresses the corresponding switching rate until the gap
approaches its branch-specific target. For the ON branch,
$a_{\rm on}=1.8$~nm, whereas for the OFF branch,
$a_{\rm off}=1.2$~nm (\Cref{tab:pickett}).

\begin{figure}[!ht]
\centering
\includegraphics[width=6.5cm,height=4cm,trim=0 0 0 4.7cm, clip]{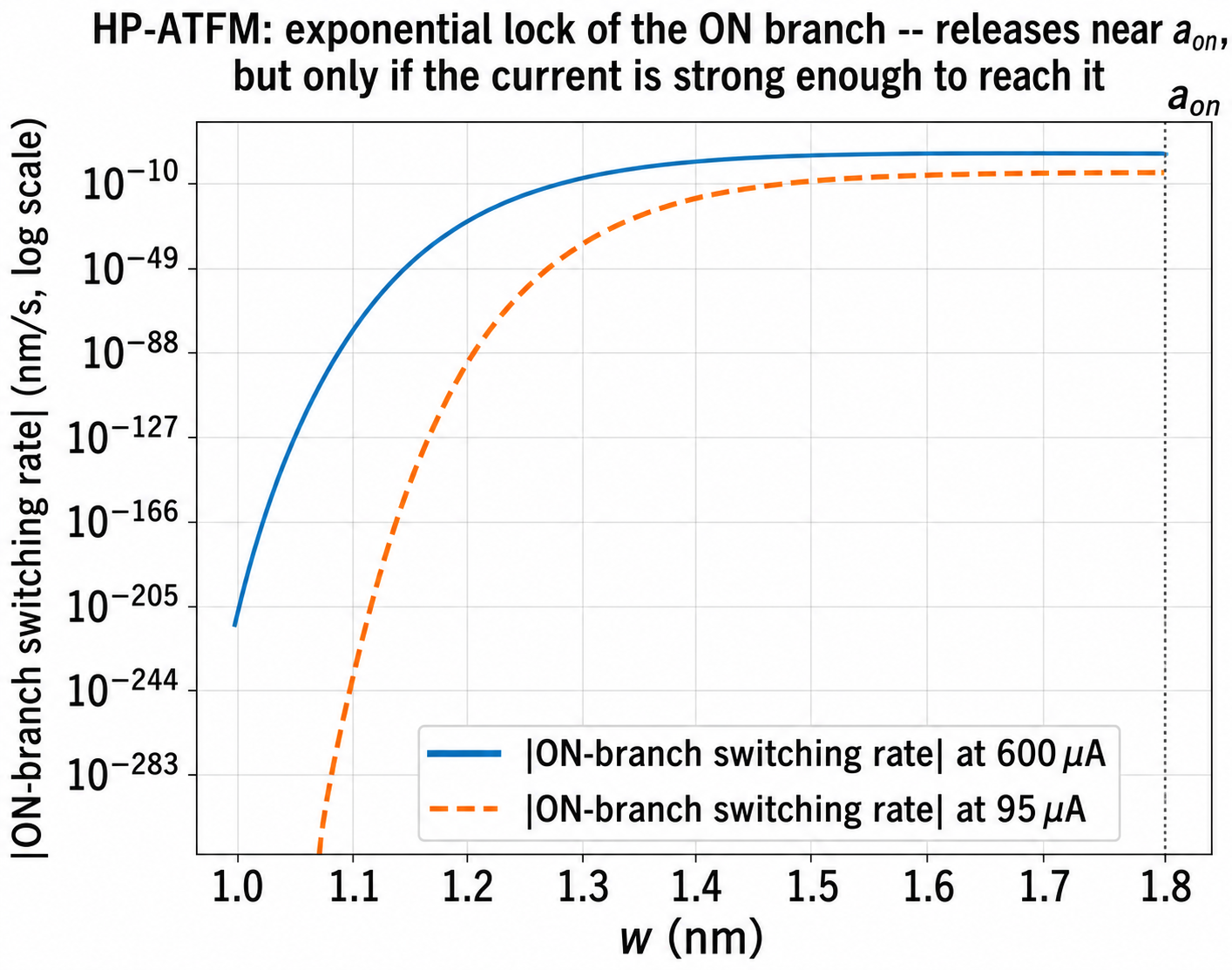}
\caption{ATFM: magnitude of the ON-branch switching rate as a
function of gap width $w$ at $I_0\approx600\,\mu$A (solid) and
$I_0\approx95\,\mu$A (dashed), on a logarithmic scale. The rate is
negligible at the lower current but increases by many orders of
magnitude near $w\approx1.4$~nm at the higher current.}
\label{fig:on-threshold}
\end{figure}

The resulting threshold
behavior is illustrated in \Cref{fig:on-threshold}: at
$I_0\approx95\,\mu$A, the ON-branch rate remains below
$10^{-12}$~nm/s over the range reached by the gap, effectively
preventing return. At $I_0\approx600\,\mu$A, the same rate increases by many orders of magnitude once $w$ exceeds approximately
$1.35$~nm, allowing the reverse switching branch to become active.

\begin{figure}[!ht]
\centering
\includegraphics[width=8.5cm,height=3.8cm,trim=0 0 0 1.95cm, clip]{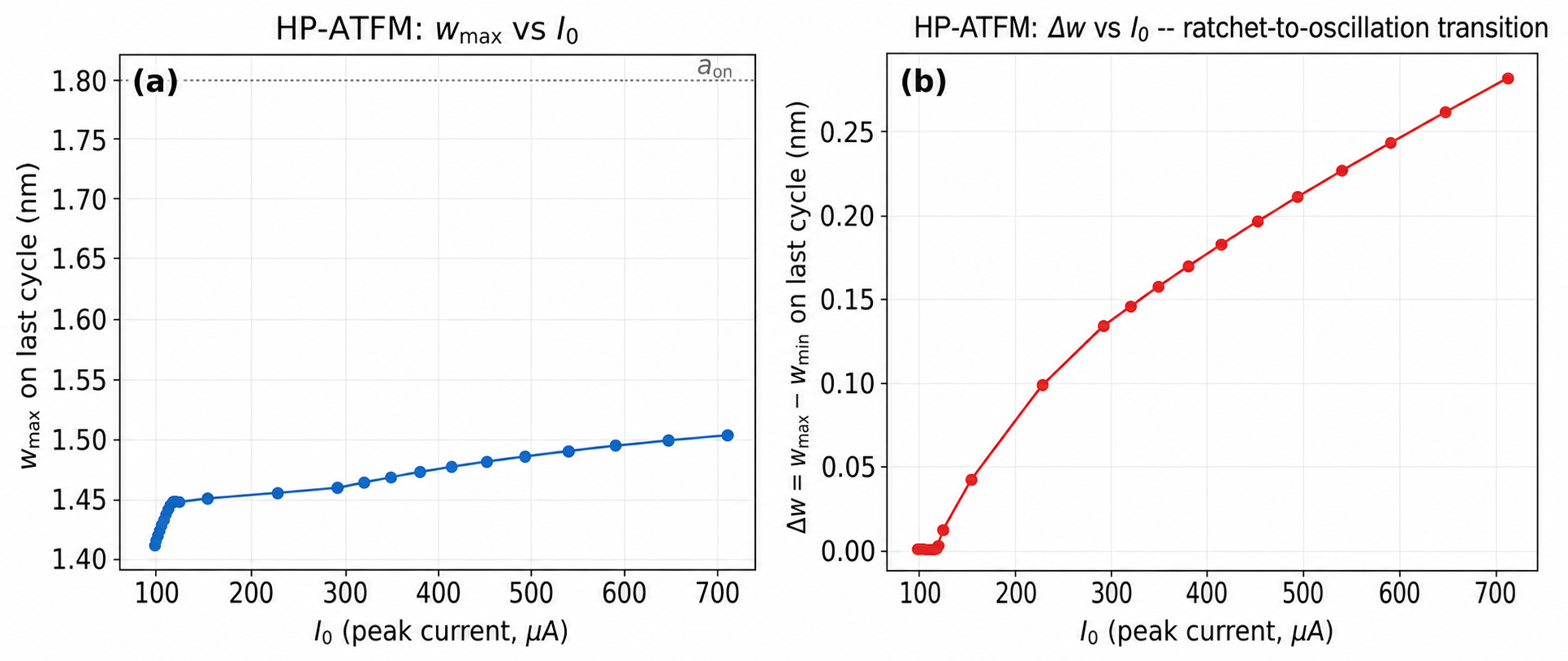}
\caption{ATFM: $w_{\max}$ (a) and
$\Delta w=w_{\max}-w_{\min}$ (b) on the last simulated cycle as a
function of peak current $I_0$, swept through the applied voltage
amplitude ($f=1$~Hz, $\Ea=0.7$~eV, 25 points,
$I_0\in[98,711]\,\mu$A).}
\label{fig:w-vs-I0}
\end{figure}

To establish that this transition is not specific to the two selected operating points, the gap excursion
\begin{equation}
\Delta w=w_{\max}-w_{\min}
\end{equation}
was evaluated over a fine sweep of voltage amplitudes corresponding
to $I_0\in[98,711]\,\mu$A (\Cref{fig:w-vs-I0}). Three regimes emerge: $\Delta w<0.002$~nm for $I_0\lesssim120\,\mu$A, a sharp increase over approximately $120$--$154\,\mu$A, and a smooth increase from $0.04$ to $0.28$~nm for $I_0\gtrsim154\,\mu$A. The transition is therefore associated with the current scale of the Pickett kinetic calibration, including $b=500\,\mu$A and $i_{\rm off}=115\,\mu$A (\Cref{tab:pickett}), rather than with an arbitrary voltage-amplitude threshold.

This threshold sensitivity follows directly from the current scale of the Pickett switching kinetics, characterized by
$i_{\rm on}$, $i_{\rm off}$, and $b$ \cite{PickettThesis}. At
currents substantially below this calibration range, ATFM enters a ratchet regime in which the gap evolves predominantly in one
direction instead of forming the bounded switching cycle observed at higher currents.

\begin{figure}[!ht]
\centering
\includegraphics[width=8.5cm,height=3.8cm,trim=0 0 0  1.85cm, clip]{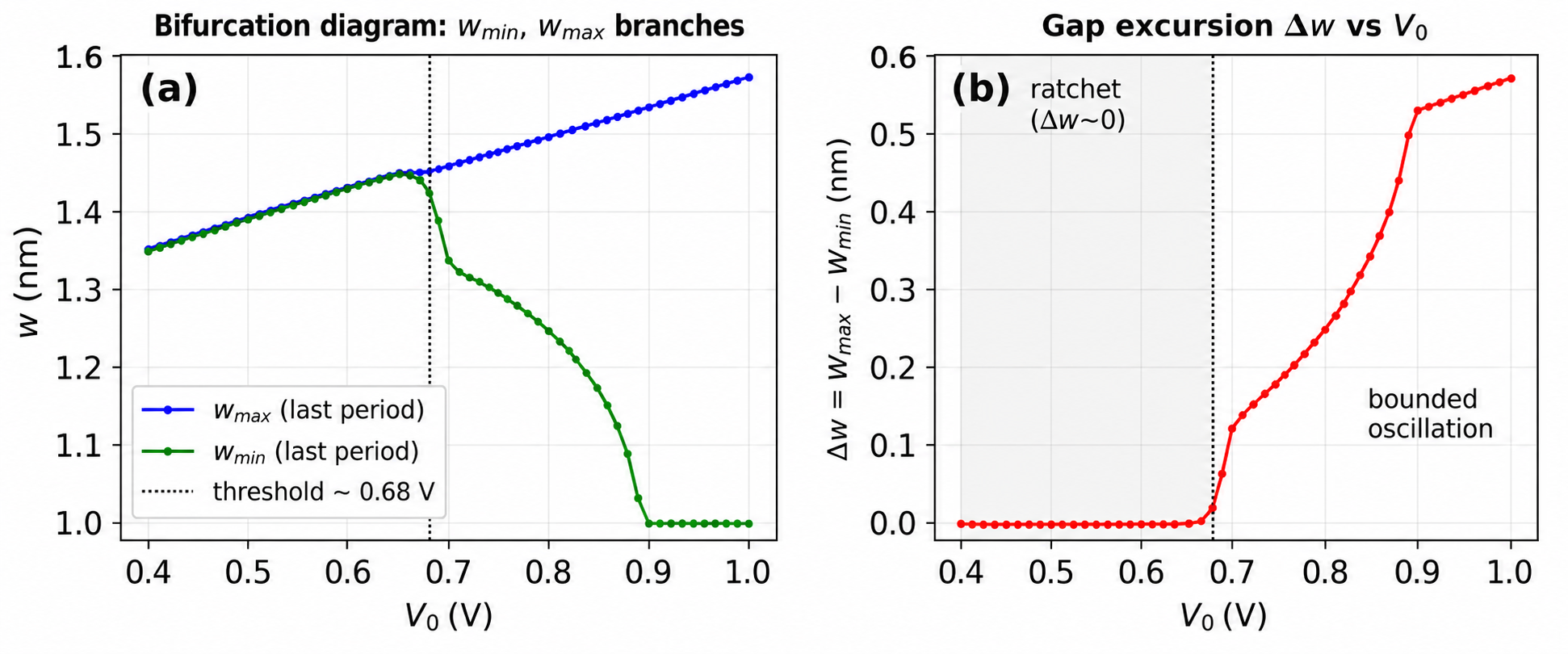}
\caption{Amplitude-sweep diagram of the ratchet-to-oscillation transition
in ATFM ($f=1$~Hz, $\Ea=0.7$~eV). (a) Last-period gap extrema $w_{\min}$ and
$w_{\max}$ versus $V_0$.
(b) Corresponding gap excursion $\Delta w=w_{\max}-w_{\min}$, which is
negligible below the threshold and rises sharply beyond it.}
\label{fig:bifurcation-V0}
\end{figure}

Sweeping the excitation amplitude $V_0$ directly provides a compact amplitude-sweep view of this transition. \Cref{fig:bifurcation-V0} reports the extrema $w_{\min}$ and $w_{\max}$ of the gap over the last simulated period as functions of $V_0$. Below a threshold $V_0^{\ast}\approx0.68$~V the two branches coincide: the gap has drifted to a period-fixed value and no longer returns within a cycle ($\Delta w\simeq0$), the signature of the ratchet regime.

Above $V_0^{\ast}$ the branches separate abruptly, and a bounded oscillation of finite amplitude $\Delta w$ emerges, growing continuously with $V_0$. The threshold coincides with the current-scale transition of \Cref{fig:w-vs-I0} and with the exponential-lock argument of \Cref{eq:gapthermal}, indicating that the ratchet-to-oscillation transition is an amplitude-controlled, threshold-like feature of the electrothermal dynamics rather than a peculiarity of a particular operating point.

\begin{figure}[!ht]
\centering
\includegraphics[width=8cm,height=4cm,trim=0 0 0 0cm, clip]{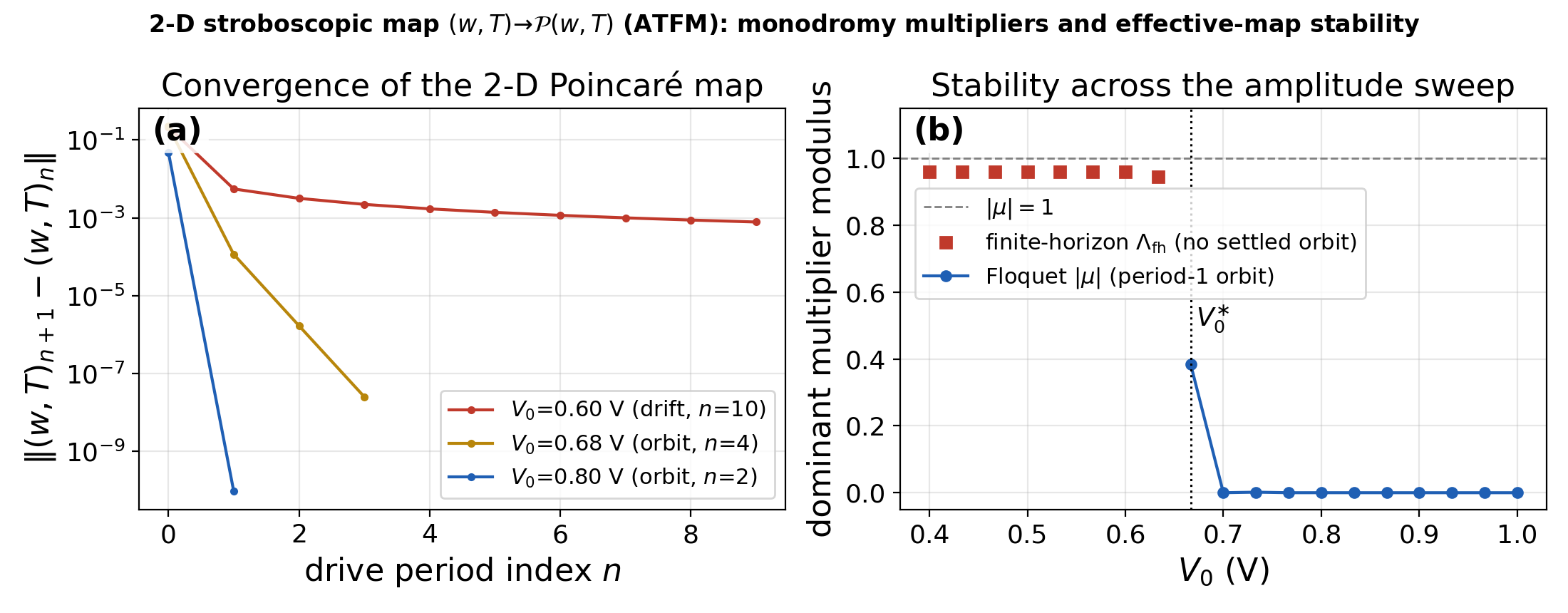}
\caption{Stroboscopic-map characterization of the
ratchet-to-oscillation transition in ATFM
($f=1$~Hz, $\Ea=0.7$~eV).
(a) Per-period state increment: above
$V_0^\ast\approx0.68$~V,  (b) Modulus of the dominant Floquet multiplier for the established period-1 orbit. It remains inside the unit circle and does not cross $\pm1$, supporting a sharp threshold-like transition.}
\label{fig:poincare-floquet}
\end{figure}

To distinguish the threshold-like transition from a classical local bifurcation, we inspect the stroboscopic map of the full state $(w,T)$, sampled once per drive period
(\Cref{fig:poincare-floquet}). Above $V_0^\ast \approx 0.68$~V, the map converges within a few periods to a stable period-1 orbit (\Cref{fig:poincare-floquet}(a)).  The full state
$(w_{n+1},T_{n+1})=\mathcal{P}(w_n,T_n)$ is sampled once per drive period and integrated using the Radau solver
($\mathrm{rtol}=10^{-7}$; $\mathrm{atol}_w=10^{-9}$~nm;
$\mathrm{atol}_T=10^{-2}$~K). The monodromy matrix
$D\mathcal{P}$ is evaluated by central finite differences with
$\delta w=10^{-6}$~nm and $\delta T=10^{-1}$~K. Its dominant
Floquet multiplier remains real and strictly inside the unit circle, decreasing rapidly toward zero as the bounded oscillation develops (\Cref{fig:poincare-floquet}(b)). Below $V_0^\ast$, the state retains a slow, non-returning ratchet drift over the observation window, so no settled period-1 orbit is identified and no Floquet interpretation is assigned. Thus, the observed change is a sharp threshold-like strengthening of the orbit contraction, associated with the exponential locking of the filament kinetics, rather than a classical local
bifurcation. The corresponding gap excursion is shown in
\Cref{fig:bifurcation-V0}.

\subsection{The Kolka Correction and an Extended Operating Window}
\label{sec:kolka-window}

\begin{figure*}[!ht]
\centering
\includegraphics[width=16cm,height=3.8cm,trim=0 0 0 0cm, clip]{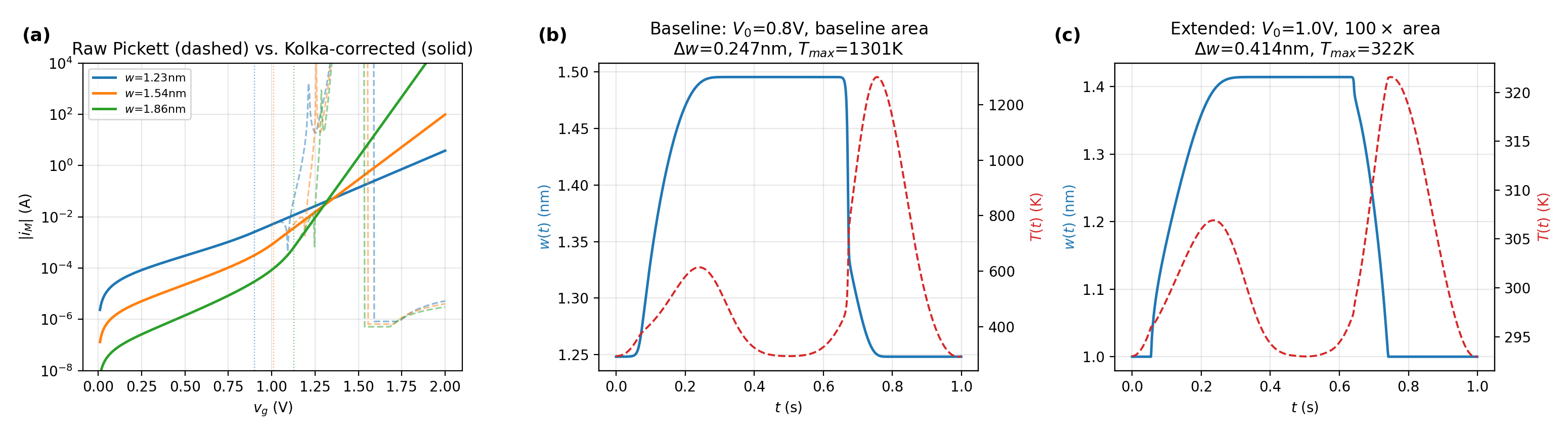}
\caption{(a) Tunneling characteristic $|I(\Vg)|$ at three gap widths,
comparing the raw Pickett relation (dashed) with the Kolka-corrected
relation (solid, \Cref{eq:kolka-extrap}); vertical dotted lines mark
$\Vg^0(w)$. The correction reproduces the raw model below threshold
and restores monotonicity above it. (b) Baseline operating point
($V_0=0.8$~V, baseline active area) with
$\Delta w=0.247$~nm and $T_{\max}=1301$~K. (c) Extended operating
point ($V_0=1.0$~V, active area enlarged by $100\times$) with
$\Delta w=0.414$~nm and $T_{\max}=322$~K. Panels (b) and (c) are
reproduced with the \texttt{SPICE} implementation in
Appendix~\ref{app:SPICE-validation}.}
\label{fig:kolka-fix}
\end{figure*}

\Cref{fig:kolka-fix} verifies the corrected tunneling relation of
\Cref{eq:kolka-extrap}. The raw characteristic becomes negative or
diverges near $\Vg\sim1$--$1.5$~V, depending on $w$, whereas the
corrected characteristic remains smooth and monotonic throughout the
tested range up to $2$~V. Below the threshold $\Vg^0(w)$, the
corrected relation reproduces the original characteristic.

At the baseline active area, the useful operating range remains
thermally constrained despite the correction: increasing $V_0$ beyond
approximately $0.85$~V drives $T_{\max}$ above $1800$~K within a
further $0.05$~V. This limit is imposed by the thermal resistance
$\Rth$ rather than by the corrected tunneling relation.

\begin{figure*}[!ht]
\centering
\includegraphics[width=17.3cm, height=4cm]{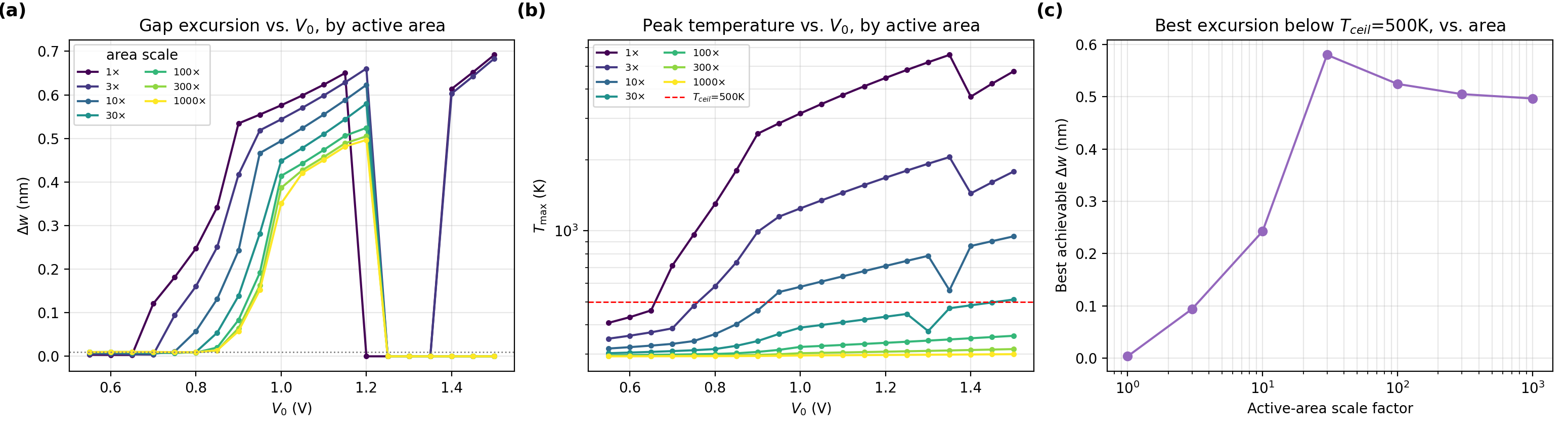}
\caption{Systematic characterization of the extended operating window
for active-area scale factors from $1\times$ to $1000\times$ baseline
($20$-point $V_0$ sweep, $\Ea=0.7$~eV). (a) Gap excursion
$\Delta w$ versus $V_0$. (b) Peak temperature $T_{\max}$ versus
$V_0$, with a $500$~K reference ceiling. (c) Maximum
$\Delta w$ achievable below this temperature ceiling as a function of
active area.}
\label{fig:area-voltage-runaway}
\end{figure*}

The dependence of this constraint on the active area is quantified in
\Cref{fig:area-voltage-runaway}. At small areas
($\lesssim10\times$ baseline), overheating (the temperature-bound
validity limit) is the dominant
limitation. At larger areas ($\gtrsim30\times$ baseline), the temperature
remains below the imposed thermal ceiling, but a second limitation
emerges: the gap becomes locked in the ON state above a kinetic
threshold near $V_0\approx1.2$--$1.3$~V. Between these regimes, the
maximum gap excursion compatible with $T_{\max}<500$~K reaches
approximately $0.58$~nm near $30\times$ the baseline area,
$V_0\approx1.2$~V, and $T_{\max}\approx432$~K. The excursion decreases
slightly at still larger areas, reaching approximately $0.50$~nm at
$1000\times$ baseline.

The active area, therefore, controls a trade-off between thermal
management and switching accessibility. Increasing $A_b$ reduces
self-heating but does not monotonically improve the switching
performance because the kinetic ON-lock eventually becomes the
limiting mechanism.

\subsection{Sensitivity of the Thermal Activation Factor to the Activation Energy}
\label{sec:gamma-sensitivity}

The thermal activation factor $\Gamma(T,\Ea)\equiv\Gamma(T)$
(\Cref{eq:thermalfactor}) follows an Arrhenius-type form commonly used
to describe thermally activated processes, including thermally activated
ionic transport and switching kinetics in resistive-switching devices.
In the present ATFM formulation, however, $\Gamma(T)$ acts as a
multiplicative modulation of the filament switching-rate prefactor
rather than directly representing an ionic drift velocity.

\begin{figure}[!ht]
\centering
\includegraphics[width=8cm,height=4.7cm,trim=0 0.5cm 0 1.5cm, clip
]{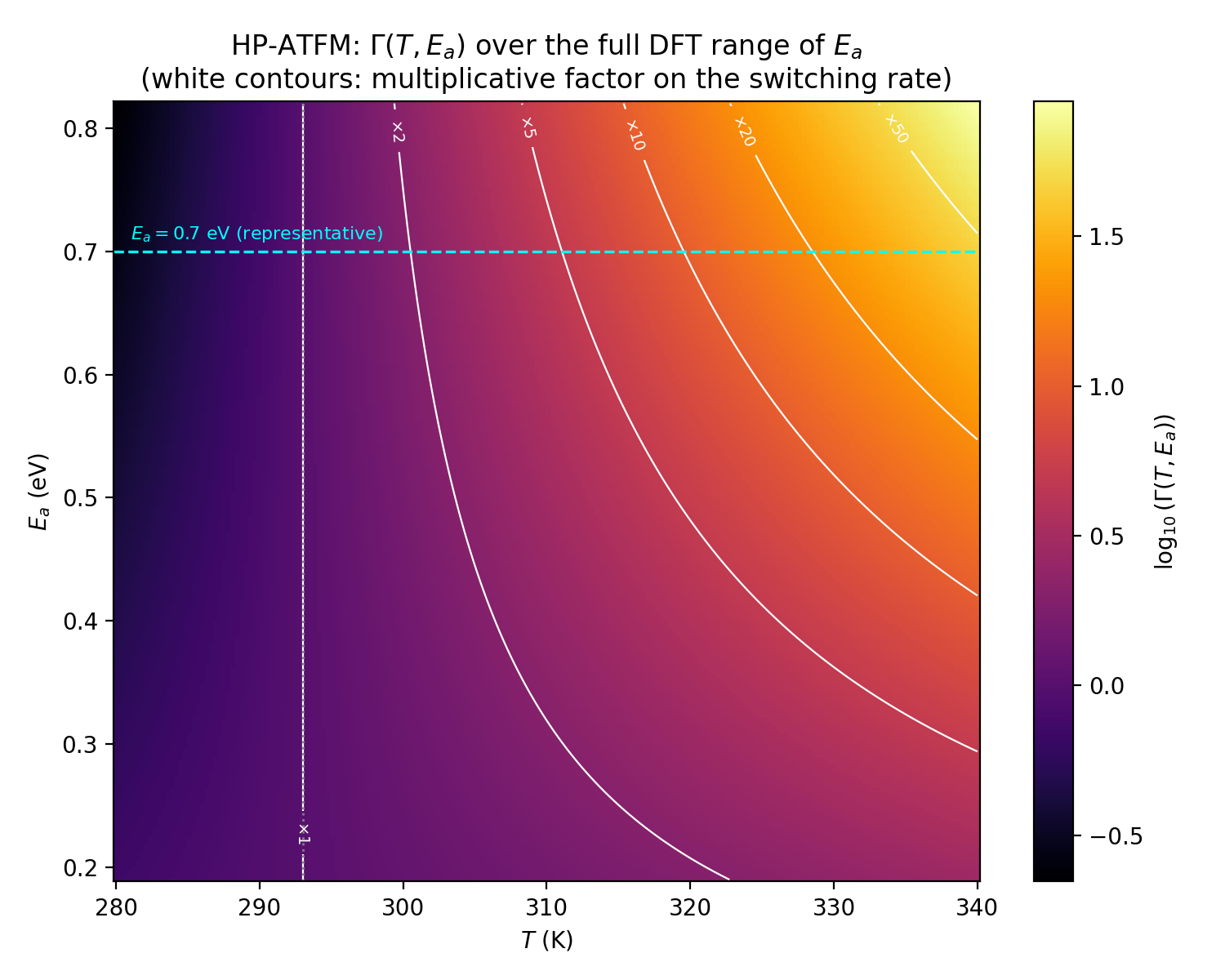}
\caption{ATFM: continuous map of $\Gamma(T,\Ea)$ on a logarithmic
scale over $\Ea\in[0.19,0.82]$~eV and $T\in[280,340]$~K. White contours
mark selected multiplicative factors ($\times1$ to $\times50$); the
cyan dashed line indicates the representative value $\Ea=0.7$~eV.
The Arrhenius factor quantifies the thermal activation of the
filament switching-rate prefactor, whose dynamical effect is governed
by the nonlinear structure of $F_{\rm gap}$.}
\label{fig:gamma-2D}
\end{figure}

\Cref{fig:gamma-2D} maps $\Gamma(T,\Ea)$ over the DFT-motivated range
$\Ea\in[0.19,0.82]$~eV and $T\in[280,340]$~K. At the representative
value $\Ea=0.7$~eV, $\Gamma$ reaches approximately $3.2$ at $306$~K
and $10.4$ at $320$~K, demonstrating the strong intrinsic thermal
activation available to the switching kinetics.

The Arrhenius dependence alone does not determine the resulting gap
dynamics. In ATFM, $\Gamma(T)$ multiplies the nonlinear function
$F_{\rm gap}(I,w)$, whose exponential lock structure controls the
current-gated transition identified in
\Cref{fig:on-threshold,fig:w-vs-I0}. Consequently, the same
multiplicative change in $\Gamma$ can have very different dynamical
effects depending on the operating point. Far below the switching
threshold, the gap remains in the ratchet regime; close to the
threshold, a change in $\Gamma$ can alter whether the reverse branch
becomes active; and in the fully engaged oscillatory regime, the
response becomes comparatively less sensitive.

Thus, \Cref{fig:gamma-2D} quantifies the intrinsic thermal activation
factor rather than providing a direct predictor of $\Delta w$. The
observed response results from the combined action of $\Gamma(T)$,
the tunneling current, and the nonlinear threshold structure of
$F_{\rm gap}$.

\subsection{Numerical Regression Test: ATFM vs. the Classical Pickett Model}
\label{sec:validation-ATFM}

The appropriate isothermal reference for ATFM is the classical
filamentary model of Pickett \textit{et al.}
\cite{Pickett2009,PickettThesis}, because ATFM is constructed from
the same physical tunneling-gap and switching framework. The
isothermal limit corresponds to
$\Gamma(T)=1$ in \Cref{eq:thermalfactor}, recovered as
$\Ea\rightarrow0$. This limit is tested numerically by comparing
ATFM with an independently simulated classical Pickett model.

\begin{figure}[!ht]
\centering
\includegraphics[width=5.5cm,height=4cm,trim=0 0 0 0.85cm, clip]{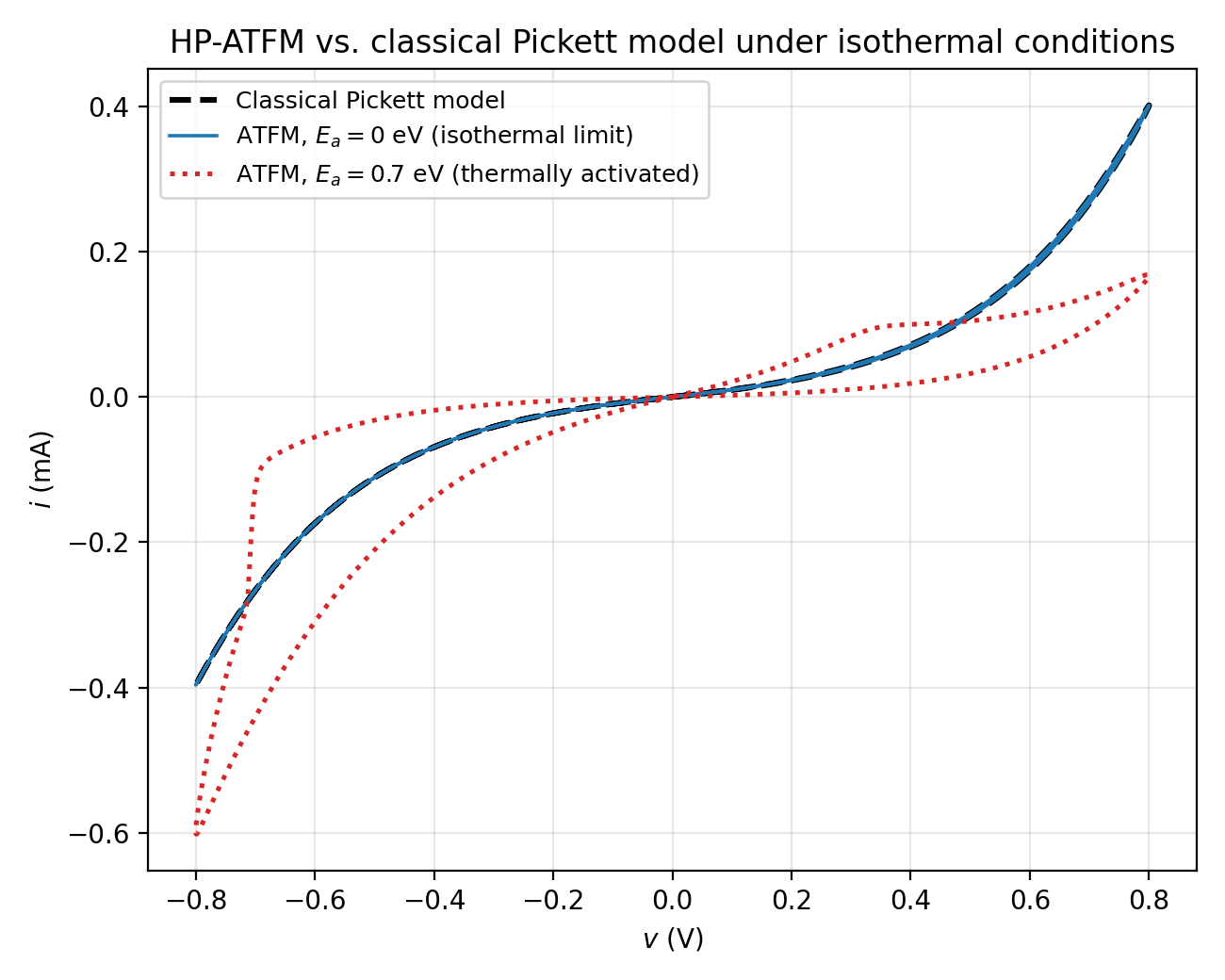}
\caption{Comparison between the classical Pickett compact model and
ATFM under isothermal conditions at $V_0=0.8$~V and $f=1$~Hz. The
classical Pickett model and ATFM at $\Ea=0$~eV coincide within
numerical tolerance, whereas ATFM at $\Ea=0.7$~eV exhibits a
substantially wider gap excursion.}
\label{fig:validationHP3b}
\end{figure}

At the reference operating point, \Cref{fig:validationHP3b} compares
the classical Pickett model, ATFM at $\Ea=0$~eV, and ATFM at
$\Ea=0.7$~eV. The classical Pickett model and the
$\Ea=0$~eV ATFM solution coincide to within
$1.4\times10^{-7}$~A, corresponding to approximately $0.03\,\%$ of
the peak current. This agreement is consistent with numerical
tolerance and confirms the implementation of the isothermal limit.

At $\Ea=0$~eV, the gap settles into a narrow limit cycle,
$w\in[1.333,1.338]$~nm, corresponding to an excursion of only
$0.005$~nm. In contrast, $\Ea=0.7$~eV produces the much wider
$w\in[1.20,1.50]$~nm excursion observed at the reference operating
point. The comparison, therefore, demonstrates that the thermally
activated factor substantially modifies the switching dynamics and
opens the large-amplitude gap excursion.

\begin{figure}[!ht]
\centering
\includegraphics[width=8.5cm,height=6.8cm,trim=0 0 0 0cm, clip]{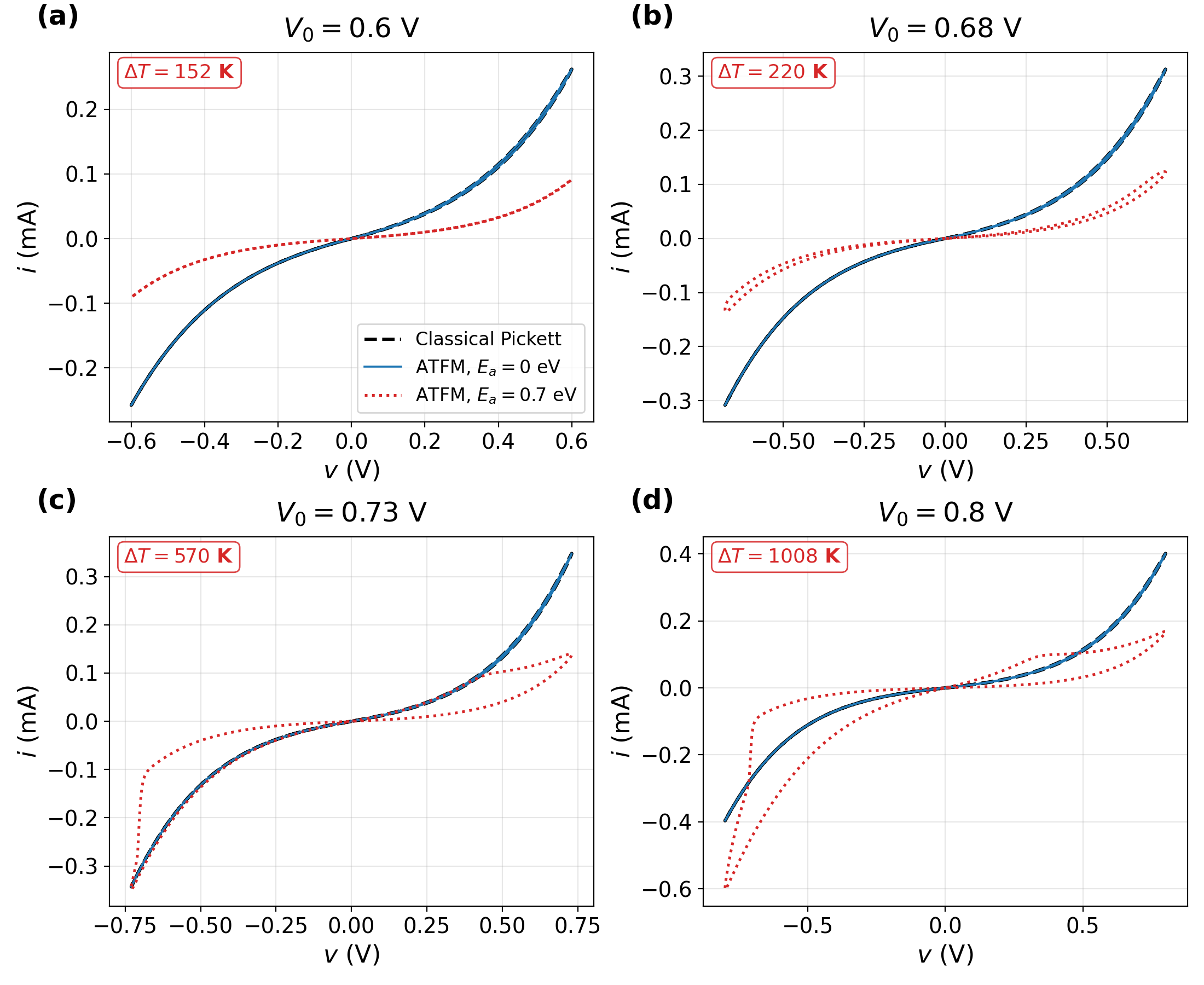}
\caption{Comparison between the classical Pickett model and ATFM
under isothermal conditions for four excitation amplitudes at
$f=1$~Hz. The classical Pickett model and ATFM at $\Ea=0$~eV
coincide within numerical tolerance, whereas $\Ea=0.7$~eV produces a
substantially wider thermally activated response.}
\label{fig:validationHP3b-grid}
\end{figure}

To test the robustness of this agreement with respect to excitation
amplitude, \Cref{fig:validationHP3b-grid} repeats the comparison for
$V_0\in\{0.60,0.68,0.73,0.80\}$~V at $f=1$~Hz. The classical Pickett
model and ATFM at $\Ea=0$~eV remain within numerical tolerance in
all cases, with deviations below $0.04\,\%$ of the peak current. The
thermally activated case, $\Ea=0.7$~eV, consistently exhibits a much
larger gap excursion, while the corresponding peak temperature rise
increases from $152$~K to $1008$~K over the considered voltage range.

\subsection{Robustness of the Isothermal Limit and Activation-Energy Dependence}
\label{sec:validation-3b-extended}

The preceding regression test varied $V_0$ at a fixed frequency.
Two additional sweeps examine the remaining independent control
parameters: excitation frequency and activation energy.

\begin{figure}[!ht]
\centering
\includegraphics[width=8.5cm,height=6.8cm,trim=0 0 0 0cm, clip]{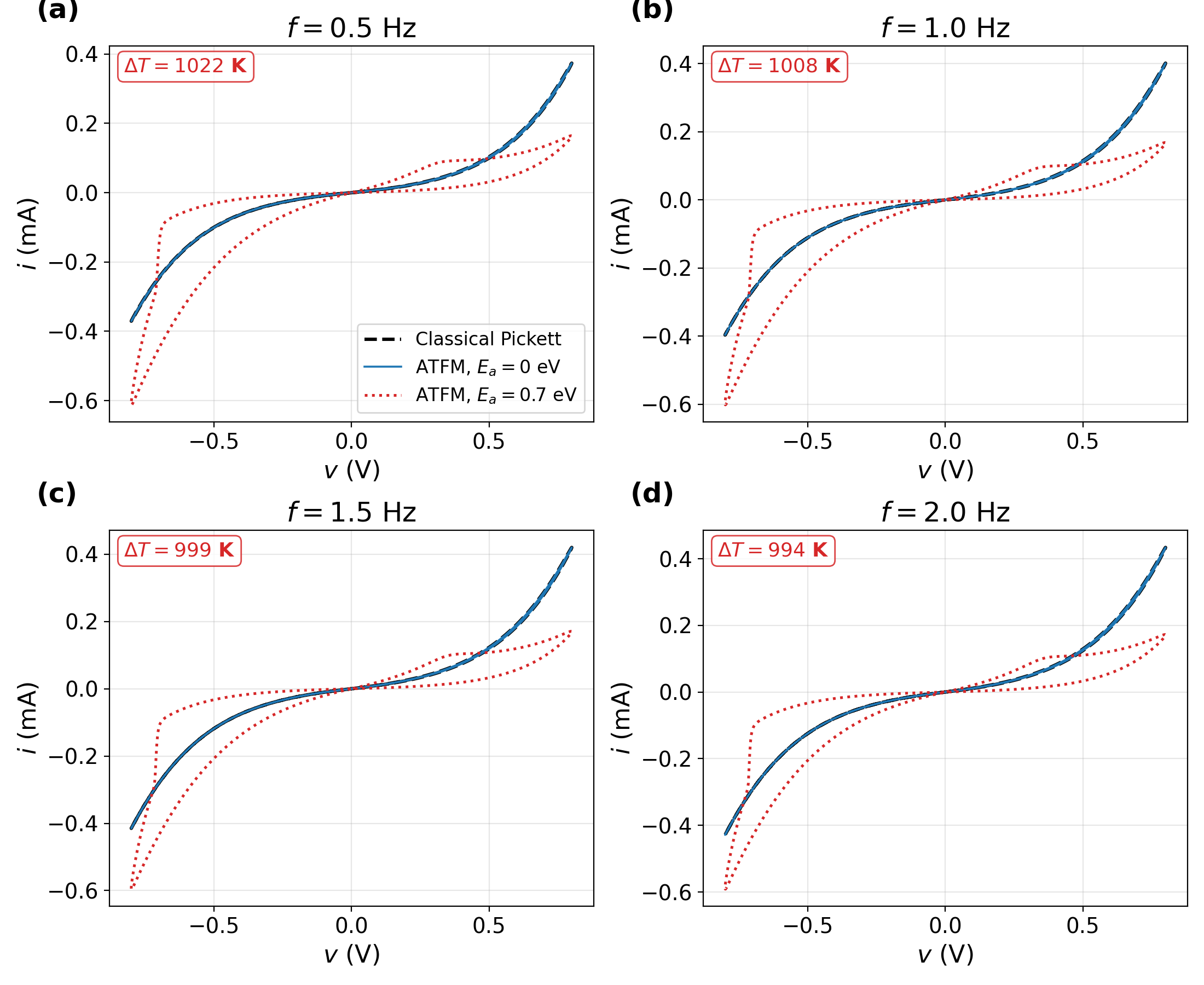}
\caption{Comparison between the classical Pickett model and ATFM
under isothermal conditions for four excitation frequencies at
$V_0=0.8$~V. The two models coincide to within $0.014\,\%$ of the peak
current in every panel.}
\label{fig:validationHP3b-freq}
\end{figure}

\Cref{fig:validationHP3b-freq} compares the classical Pickett model
with ATFM at $\Ea=0$~eV for
$f\in\{0.5,1.0,1.5,2.0\}$~Hz at $V_0=0.8$~V. The two formulations
coincide within $0.014\,\%$ of the peak current at every frequency,
confirming the isothermal limit across the tested frequency range.

For $\Ea=0.7$~eV, the peak temperature rise decreases mildly from
$1022$~K at $0.5$~Hz to $994$~K at $2.0$~Hz. Because the thermal time
constant remains many orders of magnitude shorter than all tested
periods (\Cref{sec:properties}), this weak frequency dependence is
not attributable to thermal inertia. It is instead consistent with a
dynamical phase effect: increasing the frequency reduces the time
available for $w(t)$ to approach the switching threshold during each
half-cycle, thereby slightly modifying the state and the resulting
instantaneous Joule power.

\begin{figure}[!ht]
\centering
\includegraphics[width=8.5cm,height=3.3cm,trim=0 0 0 0cm, clip]{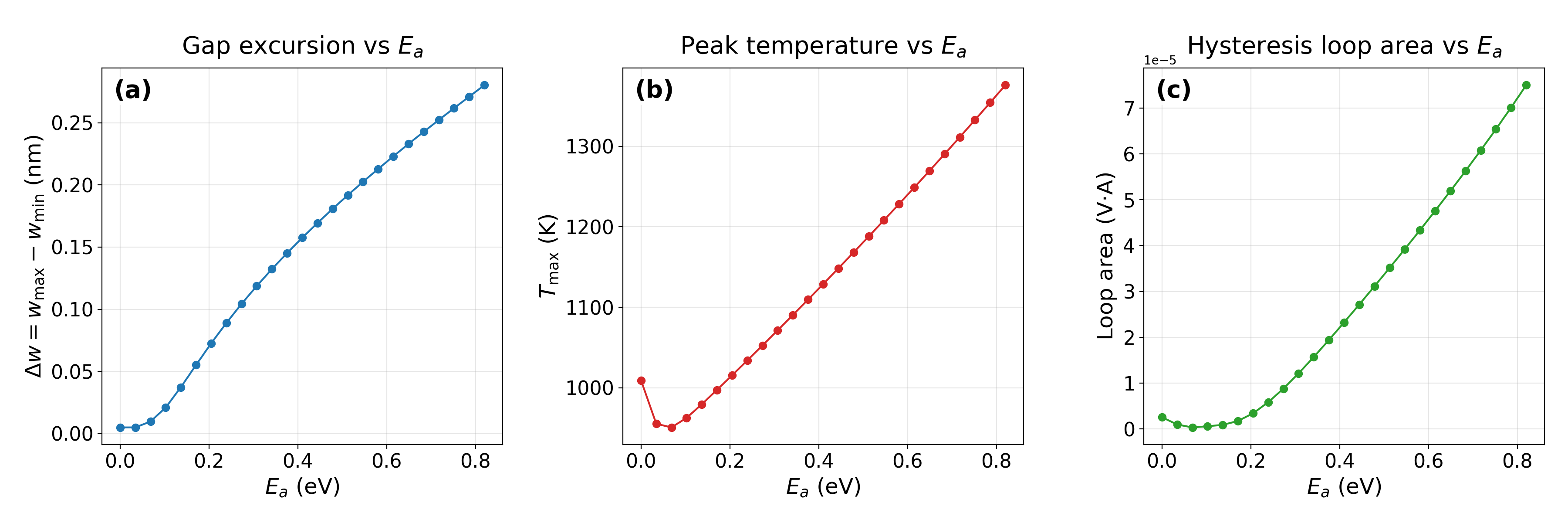}
\caption{ATFM: gap excursion $\Delta w$, peak temperature
$T_{\max}$, and hysteresis loop area as continuous functions of
$\Ea\in[0,0.82]$~eV at $V_0=0.8$~V and $f=1$~Hz.}
\label{fig:Ea-sweep-3b}
\end{figure}

The continuous activation-energy sweep in
\Cref{fig:Ea-sweep-3b} extends the analysis from the two representative
values $\Ea=0$ and $0.7$~eV to the full range
$\Ea\in[0,0.82]$~eV. The gap excursion increases smoothly from
$0.005$~nm at $\Ea=0$ to $0.28$~nm at $\Ea=0.82$~eV, without a
discontinuity across the scanned interval. The transition from the
isothermal limit to the thermally activated regime is therefore
continuous in $\Ea$, in contrast with the sharp current threshold
identified in \Cref{fig:w-vs-I0}. The peak temperature and hysteresis
loop area exhibit corresponding smooth increases, with the loop area
growing by approximately two orders of magnitude over the same
activation-energy range.

Together, the amplitude, frequency, and activation-energy sweeps
demonstrate that the isothermal limit is numerically robust, while
the thermally activated response varies smoothly with $\Ea$ and
depends strongly on the operating point.

\subsection{Temperature-Dependent Hysteresis of ATFM}
\label{sec:hysteresis-vs-Tamb-3b}

The regression tests above use the reference ambient temperature
$\Tamb=293$~K. To examine the influence of the thermal environment,
$\Tamb$ is varied while the material reference temperature
$T_0=293$~K is kept fixed. The latter is a calibration temperature
associated with the Pickett switching-rate parameters and is therefore
not identified with the operating ambient temperature.

\begin{figure}[!ht]
\centering
\includegraphics[width=8cm,height=4.3cm,trim=0.3cm 0 0 1.8cm, clip]{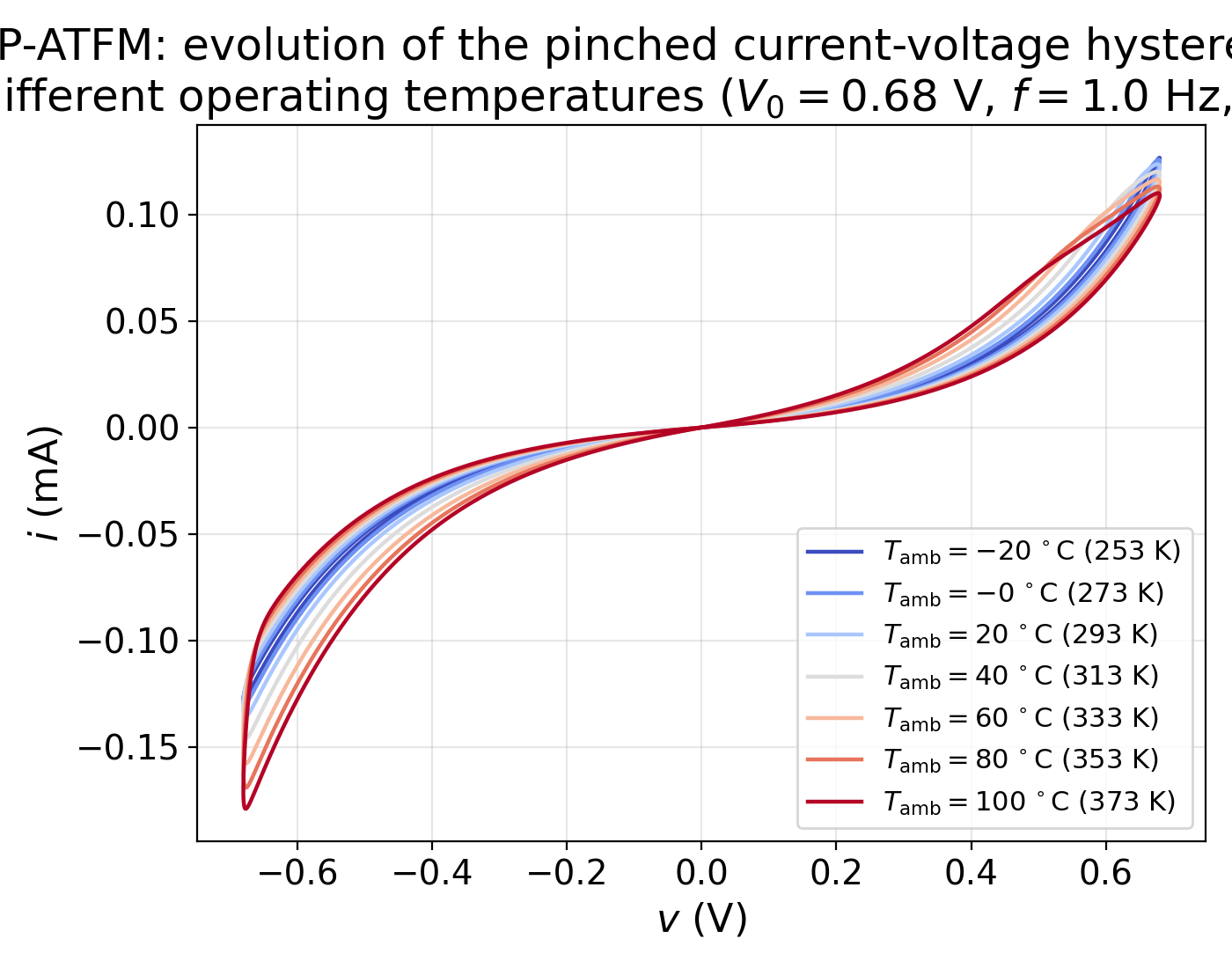}
\caption{ATFM: evolution of the pinched current--voltage hysteresis
loops with ambient temperature $\Tamb$ at $V_0=0.68$~V,
$f=1$~Hz, and $\Ea=0.7$~eV, with $T_0=293$~K fixed. The loop widens
monotonically as the ambient temperature increases from
$-20\,^\circ$C to $100\,^\circ$C.}
\label{fig:hysteresis-tamb-3b}
\end{figure}

At $V_0=0.68$~V, $f=1$~Hz, and $\Ea=0.7$~eV, the gap excursion
increases monotonically from $0.0066$~nm at
$\Tamb=-20\,^\circ$C to $0.083$~nm at
$\Tamb=100\,^\circ$C, corresponding to an increase by a factor of
approximately $13$. The response remains smooth throughout the
tested range, with the pinched current--voltage loops widening
monotonically as $\Tamb$ increases (\Cref{fig:hysteresis-tamb-3b}).

The moderate sensitivity of $\Delta w$ to $\Tamb$ reflects the fact
that the Arrhenius factor does not act as an unconstrained linear
amplification of the switching rate. Its effect remains coupled to
the nonlinear threshold structure of $F_{\rm gap}$ discussed in
\Cref{sec:hptr3b-sensitivity}.

\subsection{Joule Heating Dynamics of ATFM}
\label{sec:joule-3b}

\begin{figure}[!ht]
\centering
\includegraphics[width=8cm,height=4.3cm,trim=0 0 0 1.5cm, clip]{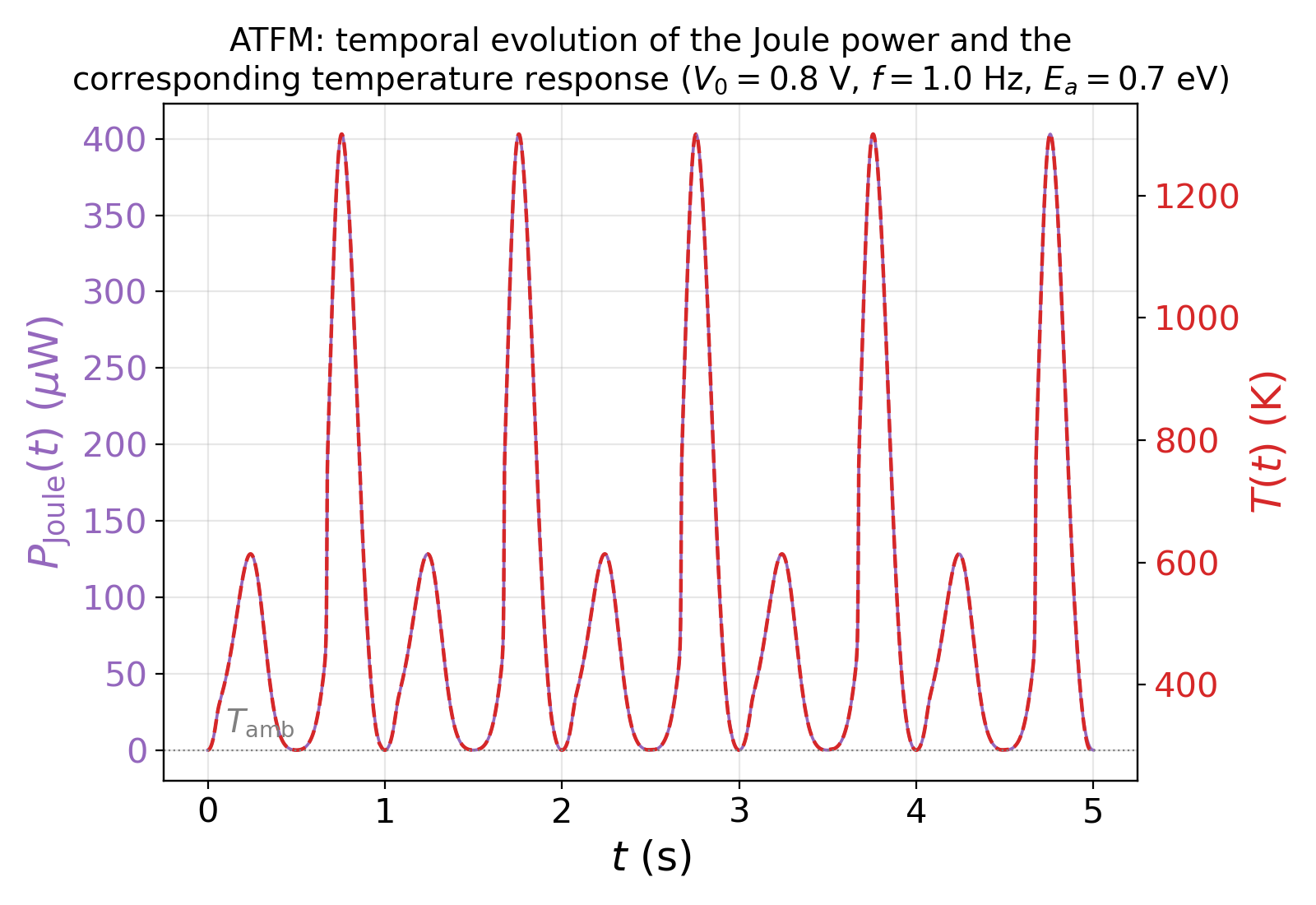}
\caption{ATFM: temporal evolution of the Joule power
$P_{\rm Joule}(t)=I(t)\Vg(t)$ and the corresponding active-region
temperature $T(t)$ at $V_0=0.8$~V, $f=1$~Hz, and $\Ea=0.7$~eV. The temperature reproduces the asymmetric structure of the Joule-power peaks, with a maximum of $1300.7$~K.}
\label{fig:joule-3b}
\end{figure}

\Cref{fig:joule-3b} resolves the relationship between the instantaneous
Joule power,
\begin{equation}
P_{\rm Joule}(t)=I(t)\Vg(t),
\end{equation}
and the temperature response at the reference operating point.

The Joule power exhibits two unequal peaks per forcing period,
approximately $128\,\mu$W and $403\,\mu$W, reflecting the asymmetric
ON/OFF switching thresholds of the filament kinetics. The temperature
response follows the same asymmetric structure and reaches
$1300.7$~K at the larger power peak before returning toward $\Tamb$.

This result provides a direct physical interpretation of the
quasi-static behavior established in
\Cref{sec:quasistatic-validation-3b}: the thermal state follows the rapidly varying dissipated power with only a very small finite lag. The Joule-power dynamics therefore provide the instantaneous thermal forcing responsible for the temperature-dependent modification of
the switching kinetics.

\subsection{Influence of the Thermal Capacitance on ATFM's Transient Dynamics}
\label{sec:Cth-3b}

The quasi-static analysis establishes the response for the
material-derived thermal time constant. To determine how the dynamics
change when thermal inertia is artificially increased,
$\Cth$ is varied while $\Rth$ and the forcing
period are kept fixed.

\begin{figure}[!ht]
\centering
\includegraphics[width=8.4cm,height=3.8cm,trim=0 0 0 0cm, clip]{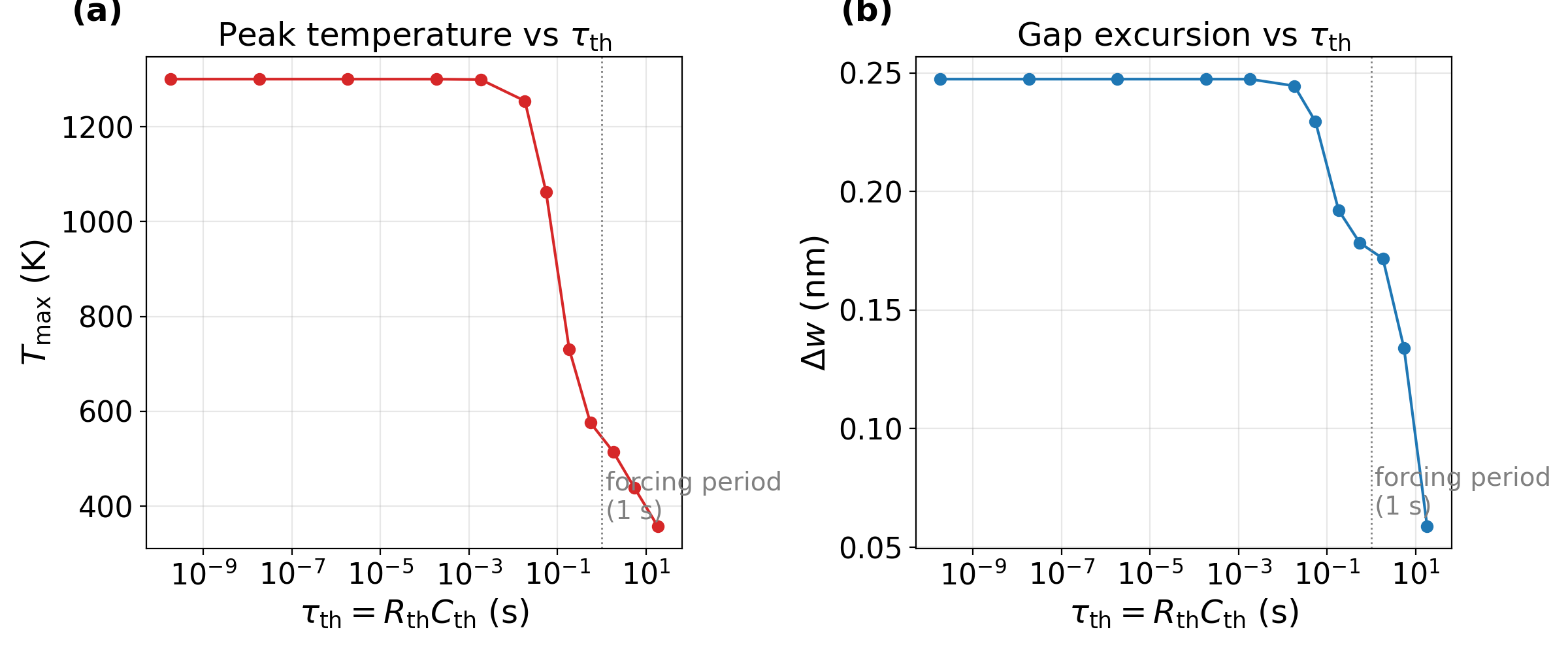}
\caption{ATFM: peak temperature $T_{\max}$ (a) and gap
excursion $\Delta w$ (b) as functions of
$\tau_{\rm th}=\Rth\Cth$, varied through
$\Cth$ at $V_0=0.8$~V, $f=1$~Hz, and $\Ea=0.7$~eV. Both
quantities remain insensitive to $\Cth$ while $\tau_{\rm th}$ is far below the forcing period and decrease once thermal inertia becomes dynamically relevant.}
\label{fig:Cth-3b}
\end{figure}

As shown in \Cref{fig:Cth-3b}, the response remains essentially
unchanged while $\tau_{\rm th}$ is much shorter than the $1$~s forcing
period. Once the thermal time constant approaches the excitation
timescale, however, the temperature spikes are strongly attenuated:
$T_{\max}$ decreases from $1300.7$~K to $357.4$~K, while the gap
excursion decreases from $0.247$ to $0.059$~nm. The reduction in
temperature weakens the thermally activated factor $\Gamma(T)$ and
thereby suppresses the switching reinforcement responsible for the
large gap excursion.

This regime requires increasing $\Cth$ by approximately
$8$--$11$ orders of magnitude above its material-derived value.
Consequently, the sweep is primarily a sensitivity test of the
quasi-static approximation and of the model response outside the
nominal parameter regime.

\subsection{Joint Sensitivity of Joule Power and Temperature to $V_0$ and $\Tamb$}
\label{sec:joule-2D-3b}

\begin{figure}[!ht]
\centering
\includegraphics[width=8.8cm,height=4cm,trim=0 0 0 0cm, clip]{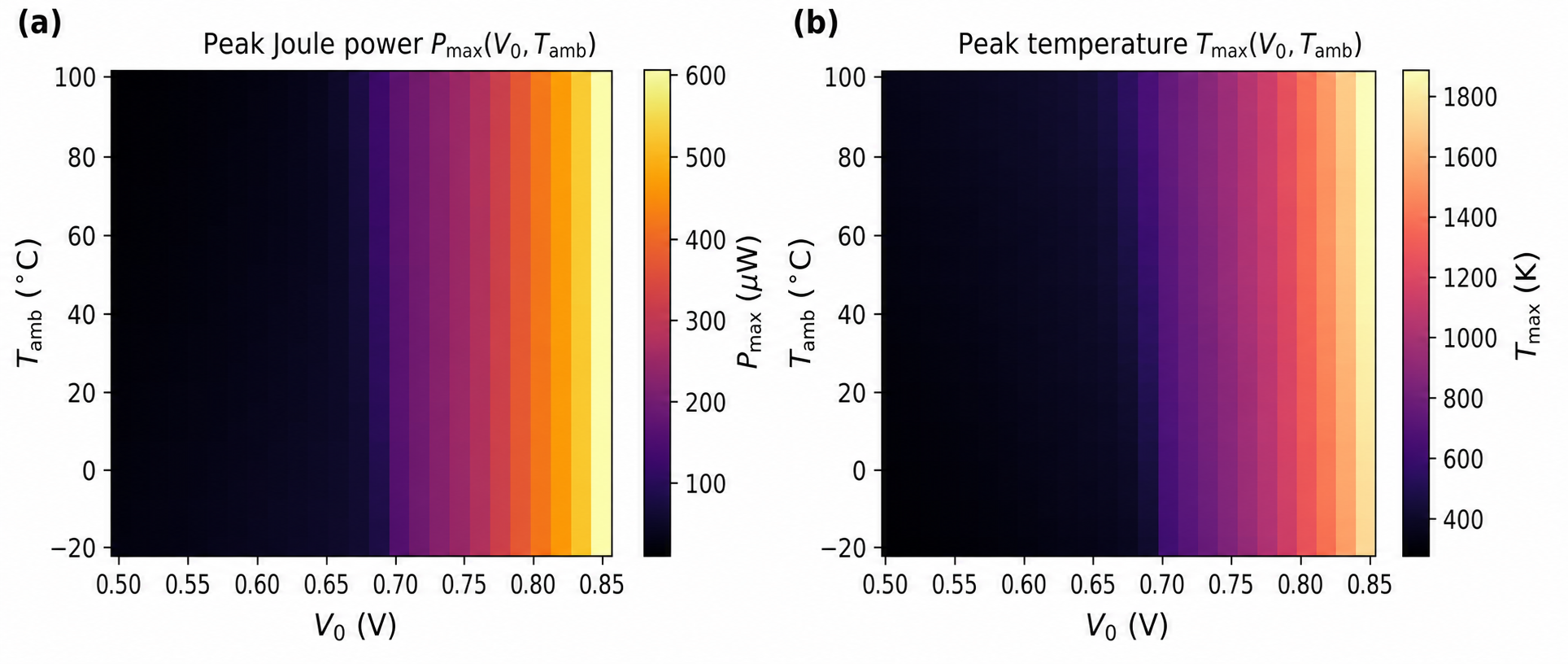}
\caption{ATFM: peak Joule power $P_{\max}(V_0,\Tamb)$ (a) and
peak temperature $T_{\max}(V_0,\Tamb)$ (b), for
$f=1$~Hz, $\Ea=0.7$~eV, and $T_0=293$~K fixed. Both quantities are
primarily controlled by $V_0$, while the effect of $\Tamb$ is
secondary.}
\label{fig:joule2D}
\end{figure}

The preceding analyses identify $V_0$ and $\Tamb$ as two important
control parameters of ATFM. Their combined influence is quantified in \Cref{fig:joule2D} by recording, for each $(V_0,\Tamb)$ pair, the peak Joule power
\[
P_{\max}=\max |I(t)\Vg(t)|
\]
and the peak temperature $
T_{\max}=\max T(t)$
over the final simulated cycle, with $T_0=293$~K fixed.

The resulting maps are dominated by the excitation amplitude.
Across the tested range, $P_{\max}$ varies by a factor of approximately
$21$ and $T_{\max}$ by a factor of approximately $5$, whereas the
effect of $\Tamb$ at fixed $V_0$ is comparatively modest:
less than $1\,\%$ for $P_{\max}$ and approximately $7\,\%$ for
$T_{\max}$ at $V_0=0.85$~V. This behavior follows from the model
structure. The ambient temperature enters the thermal balance as the
boundary condition and affects the switching kinetics through
$\Gamma(T)$, but it does not enter directly into the tunneling
current characteristic.

\begin{figure}[!ht]
\centering
\includegraphics[width=7cm,height=4.5cm,trim=0 0 0 0cm, clip]{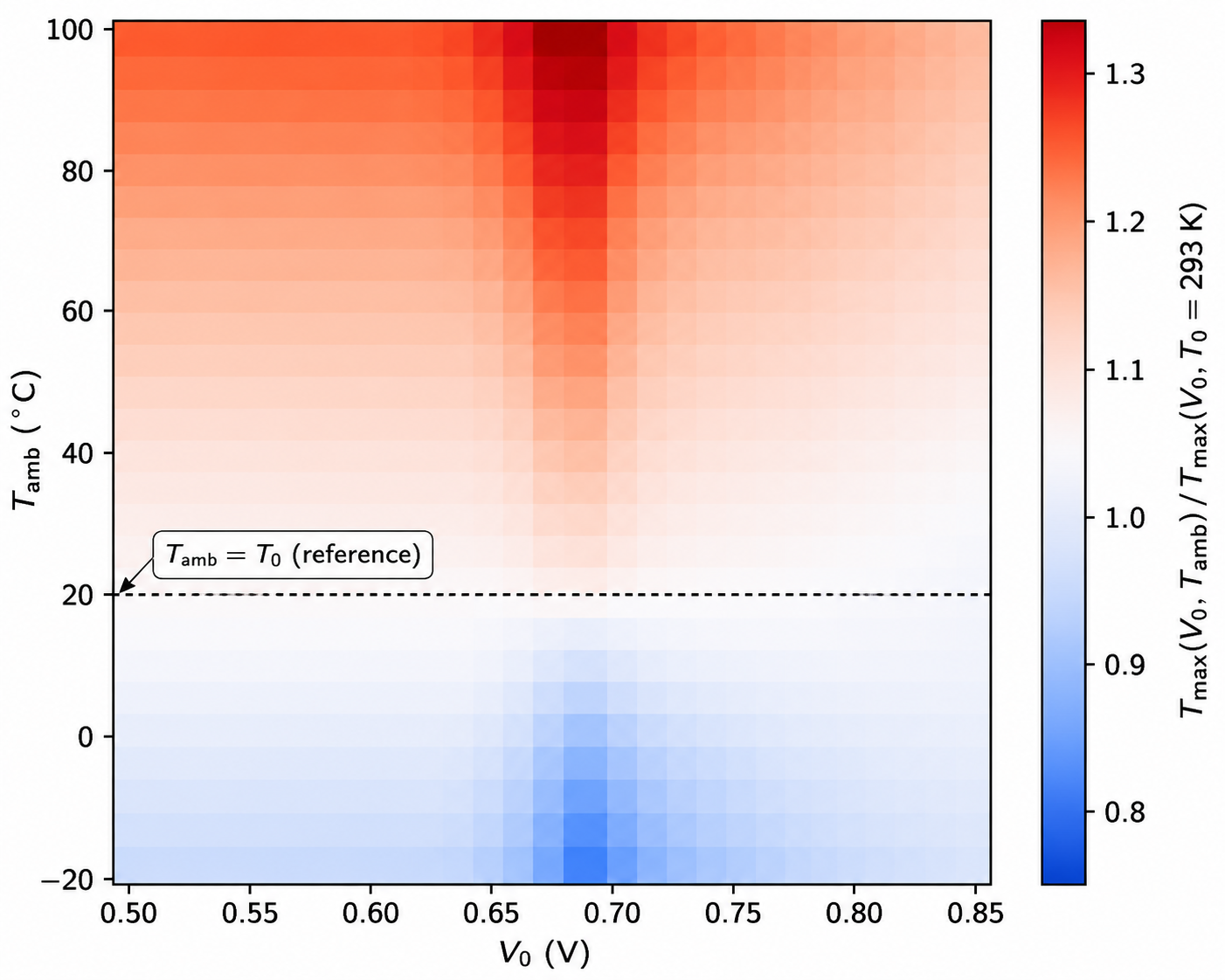}
\caption{ATFM: relative amplification of the peak temperature by
the ambient temperature,
$T_{\max}(V_0,\Tamb)/T_{\max}(V_0,T_0=293\,{\rm K})$, normalized at
fixed $V_0$.}
\label{fig:joule2D-normalized}
\end{figure}

To isolate the secondary effect of $\Tamb$, the normalized quantity
\[
\frac{T_{\max}(V_0,\Tamb)}
{T_{\max}(V_0,T_0=293\,{\rm K})}
\]
is shown in \Cref{fig:joule2D-normalized}. The relative amplification
is strongly nonuniform in $V_0$ and reaches a maximum near
$V_0\approx0.69$~V, where it reaches approximately $\times1.27$ at
$\Tamb=100\,^\circ$C, compared with approximately $\times1.05$ at
$V_0=0.85$~V. The amplification peaks near
$V_0\approx0.68$~V, coinciding with the ratchet-to-oscillation transition identified in \Cref{fig:w-vs-I0}.

A finite-difference estimate of the local voltage sensitivity further
shows that
$dT_{\max}/dV_0$ increases from approximately $417$~K/V at
$V_0=0.55$~V to $10{,}758$~K/V at $V_0=0.68$~V before decreasing to
approximately $8127$~K/V at $V_0=0.80$~V. The enhanced response near
$V_0\approx0.68$~V coincides with the ratchet-to-oscillation transition
identified in \Cref{fig:w-vs-I0}. In this region, the exponential lock
of $F_{\rm gap}$ is close to releasing, so relatively small changes in
the thermally activated switching rate can shift the system between
qualitatively different dynamical responses.

The joint map, therefore, provides independent confirmation that
the strongest sensitivity does not occur simply at the largest
$V_0$ or $\Tamb$, but near the nonlinear switching threshold where
the system changes from the ratchet regime to bounded oscillatory
dynamics.

\section{Joint Global Sensitivity Analysis}
\label{sec:sobol-3b}

The preceding analyses established that the electrothermal response is
strongly controlled by the excitation amplitude and that the ambient
temperature can produce a localized response amplification near the
ratchet-to-oscillation threshold. However, these analyses do not
separate the direct effect of each parameter from its contribution
through parameter interactions. We therefore perform a variance-based
global sensitivity analysis using the Sobol' method
\cite{Sobol2001} with Saltelli sampling \cite{Saltelli2010}. The
analysis is first carried out for $(V_0,\Tamb,\Ea)$ and subsequently
extended to include the thermal parameters $\Rth$ and $\Cth$.

\subsection{Method and Sampling Strategy}
\label{sec:sobol-method-3b}

Two independent quasi-random matrices
$A,B\in[0,1]^{N\times D}$ are generated using Sobol' low-discrepancy
sequences, with $N=1024$ base samples. For the initial analysis,
$D=3$ and the uncertain parameters are sampled over
$
V_0\in[0.55,0.85]~\mathrm{V},\quad
\Tamb\in[253,373]~\mathrm{K},\quad
\Ea\in[0.19,0.82]~\mathrm{eV}.
$
For each parameter $i$, a hybrid matrix $AB_i$ is constructed by
replacing the $i$th column of $A$ with the corresponding column of
$B$. The complete electrothermal model, defined by Eqs.~(6), (8), and
(9), is then evaluated for $A$, $B$, and all three hybrid matrices,
giving
\[
N(D+2)=5120
\]
model evaluations.

Each realization is integrated using the implicit Radau scheme together
with the nested root-finding procedure described in
\Cref{sec:properties}. All $5120$ simulations converge successfully.
The first- and total-order Sobol' indices are estimated as
\begin{equation}
\begin{split}
S_i &=
\frac{
\frac{1}{N}\sum_{j=1}^{N}
y_B(j)\left[y_{AB_i}(j)-y_A(j)\right]
}{
\operatorname{Var}(y)
},\\
S_{T_i} &=
\frac{
\frac{1}{2N}\sum_{j=1}^{N}
\left[y_A(j)-y_{AB_i}(j)\right]^2
}{
\operatorname{Var}(y)
},
\end{split}
\label{eq:sobol-formulas-3b}
\end{equation}
where $\operatorname{Var}(y)$ is computed from the pooled sample
$\{y_A,y_B\}$. The difference
$S_{T_i}-S_i$ measures the total contribution associated with
interactions involving parameter $i$; it does not identify a specific
pairwise interaction.

Four scalar outputs are extracted from the final simulated period at
$f=1$~Hz:
\[
\Delta w=w_{\max}-w_{\min},\quad
T_{\max},\quad
\Delta T_{\max}=T_{\max}-\Tamb,
\]
and the pinched-loop area
\[
A_{\mathrm{hyst}}=\tfrac{1}{2}\big|\oint\!\big(I\,\mathrm{d}V-V\,\mathrm{d}I\big)\big|
\]
The final period is used to remove the initial transient; the last two
periods agree to four significant figures, including near the
sensitive threshold.

The sampled voltage range is chosen to retain the physically relevant
transition while avoiding regimes that provide little additional
information. Below approximately $0.55$~V, the gap excursion is nearly
degenerate in the deep ratchet regime, whereas above approximately
$0.85$~V the baseline geometry approaches the overheating limit
identified in \Cref{sec:kolka-window}. The ratchet-to-oscillation
transition near $V_0\approx0.70$~V therefore remains inside the
sampling domain.

\subsection{Global Sensitivity to $V_0$, $\Tamb$, and $\Ea$}
\label{sec:sobol-results-3b}

The resulting Sobol' indices are summarized in
\Cref{tab:sobol-3b}. A clear hierarchy emerges: $V_0$ is the dominant
control parameter across all four outputs, with total-order indices
between $0.844$ and $0.984$. Its influence is especially strong for
the thermal observables, showing that their global variance is
primarily amplitude-driven.

The activation energy $\Ea$ is the second important contributor,
particularly for the nonlinear switching observables $\Delta w$ and
$A_{\mathrm{hyst}}$. By contrast, $\Tamb$ contributes only weakly to
the global variance over the sampled range, with total-order indices
below $0.011$. This small global contribution does not contradict the
localized temperature sensitivity identified previously: the latter
is confined to the neighborhood of the switching threshold, whereas
the Sobol' analysis measures variance over the entire parameter
domain.

The difference between first- and total-order indices further
distinguishes the thermal and switching responses. The thermal
observables are dominated by direct amplitude effects, with only weak
interaction contributions. In contrast, both $\Delta w$ and
$A_{\mathrm{hyst}}$ exhibit appreciable parameter coupling. For the
hysteresis-loop area, the interaction contributions reach $0.191$ for
$V_0$ and $0.221$ for $\Ea$, making $A_{\mathrm{hyst}}$ the output most
sensitive to nonlinear parameter interactions.

\begin{table}[!t]
\centering
\caption{First-order ($S_i$) and total-order ($S_{T_i}$) Sobol' sensitivity
indices for the ATFM model with respect to the excitation amplitude
$V_0$, ambient temperature $T_{\mathrm{amb}}$, and activation energy
$\Ea$. The quantity $S_{T_i}-S_i$ measures the contribution associated
with parameter interactions.}
\label{tab:sobol-3b}
\renewcommand{\arraystretch}{1.08}
\setlength{\tabcolsep}{4.5pt}
\begin{tabular}{@{}llccc@{}}
\toprule
Output & Parameter & $S_i$ & $S_{T_i}$ & $S_{T_i}-S_i$ \\
\midrule
\multirow{3}{*}{$\Delta w$}
& $V_0$              & 0.779 & 0.875 & 0.096 \\
& $T_{\mathrm{amb}}$ & 0.001 & 0.004 & 0.003 \\
& $\Ea$              & 0.120 & 0.220 & 0.100 \\
\midrule
\multirow{3}{*}{$T_{\max}$}
& $V_0$              & 0.943 & 0.976 & 0.033 \\
& $T_{\mathrm{amb}}$ & 0.008 & 0.011 & 0.003 \\
& $\Ea$              & 0.016 & 0.047 & 0.031 \\
\midrule
\multirow{3}{*}{$\Delta T_{\max}$}
& $V_0$              & 0.951 & 0.984 & 0.033 \\
& $T_{\mathrm{amb}}$ & -0.001 & 0.002 & 0.003 \\
& $\Ea$              & 0.015 & 0.048 & 0.032 \\
\midrule
\multirow{3}{*}{$A_{\mathrm{hyst}}$}
& $V_0$              & 0.653 & 0.844 & 0.191 \\
& $T_{\mathrm{amb}}$ & 0.003 & 0.011 & 0.008 \\
& $\Ea$              & 0.125 & 0.346 & 0.221 \\
\bottomrule
\end{tabular}

\vspace{1.5mm}
\begin{minipage}{0.96\linewidth}
\footnotesize
\textit{Notes:}
The four scalar outputs are extracted from the final simulated
period at $f=1$~Hz. They are the gap excursion
$\Delta w=w_{\max}-w_{\min}$, the peak temperature $T_{\max}$,
the peak self-heating $\Delta T_{\max}=T_{\max}-\Tamb$, and the
pinched-loop area
$A_{\mathrm{hyst}}=\tfrac{1}{2}\big|\oint\!\big(I\,\mathrm{d}V-V\,\mathrm{d}I\big)\big|$
(in V$\cdot$A).
\end{minipage}
\end{table}

\begin{figure*}[!ht]
\centering
\includegraphics[width=0.98\textwidth]{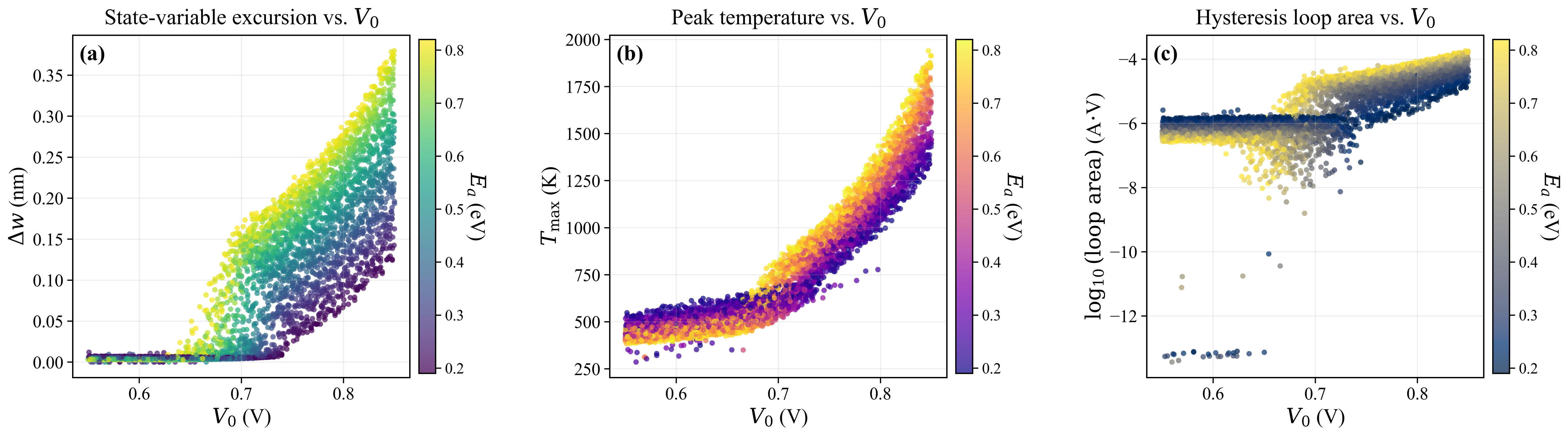}
\caption{ATFM parameter-response maps obtained from the $N=1024$
Saltelli base-sample campaign for $(V_0,\Tamb,\Ea)$, corresponding to
$5120$ model evaluations. The points are colored by $\Ea$.
(a) Gap excursion $\Delta w$. (b) Peak temperature $T_{\max}$.
(c) Hysteresis-loop area on a logarithmic scale. The maps highlight
the transition near $V_0\approx0.65$--$0.70$~V and the increasing
role of $\Ea$ in the nonlinear switching observables.}
\label{fig:sobol-maps-3b}
\end{figure*}

The parameter-response maps in \Cref{fig:sobol-maps-3b} provide the
spatial counterpart of this variance-based ranking. The gap excursion
remains nearly zero below $V_0\simeq0.65$~V and then increases rapidly,
with a progressively stronger dependence on $\Ea$. The temperature
response remains predominantly amplitude-driven, while the
activation-energy dependence becomes more visible at larger
excitation amplitudes. The hysteresis-loop area undergoes its strongest
reorganization in the same transition region before increasing with
$V_0$.

Together, the indices and maps identify
$V_0\approx0.65$--$0.70$~V as the principal transition region.
The excitation amplitude sets the location and strength of the
transition, while $\Ea$ increasingly modulates the switching response
through parameter coupling. The global indices therefore quantify the
same amplitude--kinetic coupling suggested by the preceding local
analyses, without assigning the observed interaction to a specific
second-order parameter pair.

\subsection{Extended Sensitivity Analysis with \texorpdfstring{$\Rth$}{Rth}
and \texorpdfstring{$\Cth$}{Cth}}
\label{sec:sobol-rth-3b}

The preceding analysis treats $\Rth$ and $\Cth$ as fixed
geometry-derived quantities (\Cref{tab:geometry}). We next include
both as independent multiplicative factors over
$[0.5,2]$ times their baseline values, giving a five-parameter analysis.

Varying $\Rth$ changes the thermal safety boundary. At the worst tested
combination, $\Rth$ scale $=2$ and $V_0=0.85$~V, the model reaches
$T_{\max}=3448$~K (above the rutile melting point, $\approx2116$~K), beyond its physical validity range. The upper
voltage limit is therefore reduced to $V_0=0.76$~V, which maintains
$T_{\max}<1850$~K at $\Rth$ scale $=2$. The lower voltage limit and
the ranges of $\Tamb$ and $\Ea$ remain unchanged. With $N=1024$ and
$D=5$, the analysis comprises
\[
N(D+2)=7168
\]
model evaluations, all of which converge.

The extended indices in \Cref{tab:sobol-5param-3b} confirm the dominant
role of $V_0$, whose total-order index remains above $0.75$ for every
output. The new feature is the emergence of $\Rth$ as an important
thermal parameter. Its total-order indices reach $0.34684$ for
$T_{\max}$ and $0.35229$ for $\Delta T_{\max}$, making it the second
most influential input for both thermal observables.

The influence of $\Rth$ on the switching observables is smaller but
is mainly interaction-mediated. Its total interaction contribution
reaches $0.08414$ for $\Delta w$ and $0.10622$ for
$A_{\mathrm{hyst}}$. This behavior follows directly from the
electrothermal pathway
\[
\Rth \rightarrow T \rightarrow \Gamma(T)
\rightarrow \text{gap kinetics},
\]
which allows a thermal-calibration parameter to influence switching
without appearing directly in the tunneling-current relation.

The activation energy remains important for the nonlinear switching
response, with total-order indices of $0.38681$ for $\Delta w$ and
$0.46158$ for $A_{\mathrm{hyst}}$. Its relatively large interaction
contribution confirms that the Arrhenius kinetics is most influential
when coupled to the electrical and thermal state of the device.

By contrast, $\Tamb$ remains weak over the five-parameter domain, while
$\Cth$ contributes very little to the output variance, with all
total-order indices below $0.015$. This behavior is consistent with
the short thermal time-scale identified in \Cref{sec:properties} and
with the quasi-static thermal response established in
\Cref{sec:quasistatic-validation-3b}.

\begin{table}[!t]
\centering
\caption{First-order ($S_i$) and total-order ($S_{T_i}$) Sobol' sensitivity
indices for the five-parameter ATFM model. The quantity
$S_{T_i}-S_i$ measures the contribution associated with parameter
interactions.}
\label{tab:sobol-5param-3b}
\renewcommand{\arraystretch}{1.06}
\setlength{\tabcolsep}{3.8pt}
\begin{tabular}{@{}llccc@{}}
\toprule
\textbf{Output} & \textbf{Parameter} &
$S_i$ & $S_{T_i}$ & $S_{T_i}-S_i$ \\
\midrule
\multirow{5}{*}{$\Delta w$}
& $V_0$ & 0.57539 & 0.81340 & 0.23801 \\
& $T_{\mathrm{amb}}$ & 0.00385 & 0.01289 & 0.00904 \\
& $\Ea$ & 0.17255 & 0.38681 & 0.21425 \\
& $R_{\mathrm{th}}$ & 0.01524 & 0.09938 & 0.08414 \\
& $C_{\mathrm{th}}$ & -0.00048 & 0.00009 & 0.00056 \\
\midrule
\multirow{5}{*}{$T_{\max}$}
& $V_0$ & 0.57259 & 0.75865 & 0.18606 \\
& $T_{\mathrm{amb}}$ & 0.00028 & 0.02152 & 0.02124 \\
& $\Ea$ & 0.02761 & 0.09134 & 0.06373 \\
& $R_{\mathrm{th}}$ & 0.18850 & 0.34684 & 0.15834 \\
& $C_{\mathrm{th}}$ & -0.00089 & 0.00357 & 0.00446 \\
\midrule
\multirow{5}{*}{$\Delta T_{\max}$}
& $V_0$ & 0.58903 & 0.77056 & 0.18153 \\
& $T_{\mathrm{amb}}$ & -0.00019 & 0.00600 & 0.00620 \\
& $\Ea$ & 0.02536 & 0.09277 & 0.06741 \\
& $R_{\mathrm{th}}$ & 0.19681 & 0.35229 & 0.15548 \\
& $C_{\mathrm{th}}$ & -0.00547 & 0.00363 & 0.00910 \\
\midrule
\multirow{5}{*}{$A_{\mathrm{hyst}}$}
& $V_0$ & 0.49373 & 0.82686 & 0.33313 \\
& $T_{\mathrm{amb}}$ & 0.00585 & 0.03641 & 0.03057 \\
& $\Ea$ & 0.15503 & 0.46158 & 0.30655 \\
& $R_{\mathrm{th}}$ & 0.00939 & 0.11561 & 0.10622 \\
& $C_{\mathrm{th}}$ & -0.00021 & 0.01461 & 0.01482 \\
\bottomrule
\end{tabular}
\end{table}

\begin{figure*}[!ht]
\centering
\includegraphics[width=0.95\textwidth]{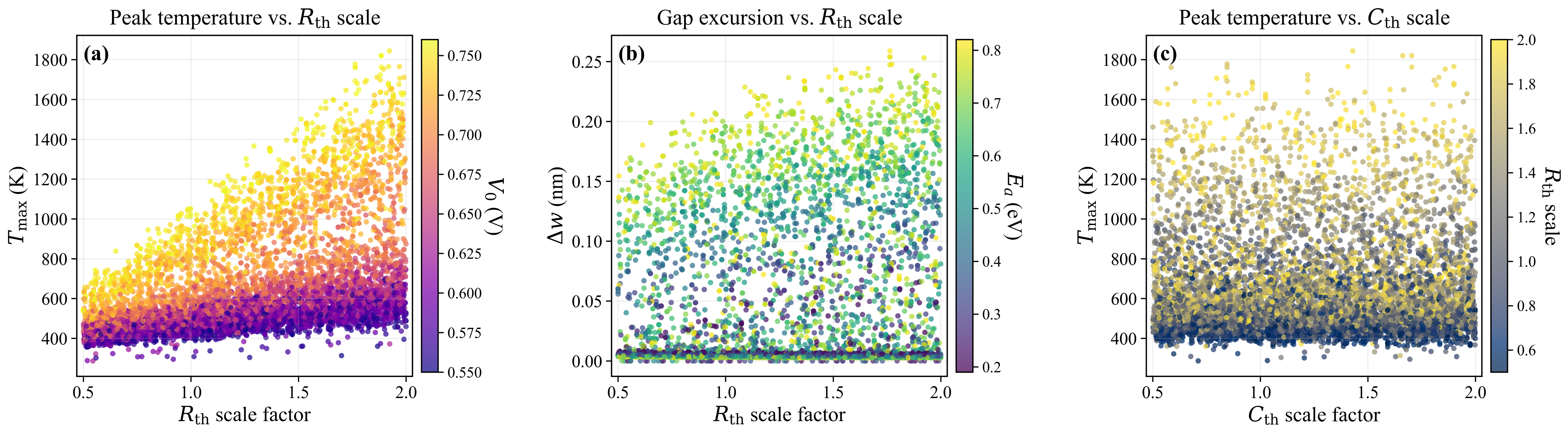}
\caption{Five-parameter ATFM Saltelli sample ($7168$ evaluations). (a) Peak temperature $T_{\max}$ versus $\Rth$ scale, colored by $V_0$. (b) Gap excursion $\Delta w$ versus $\Rth$ scale, colored by $\Ea$. (c) Peak temperature $T_{\max}$ versus $\Cth$ scale, colored by $\Rth$ scale. The maps highlight the strong thermal influence of $\Rth$ and the weak dependence on $\Cth$.}
\label{fig:sobol-5param}
\end{figure*}

\subsection{Estimation Uncertainty and Convergence}
\label{sec:sobol-uncertainty}

\begin{figure}[!t]
  \centering
  \includegraphics[width=8.3cm, height=7.5cm,
  trim=0 0 0 0.85cm,clip]{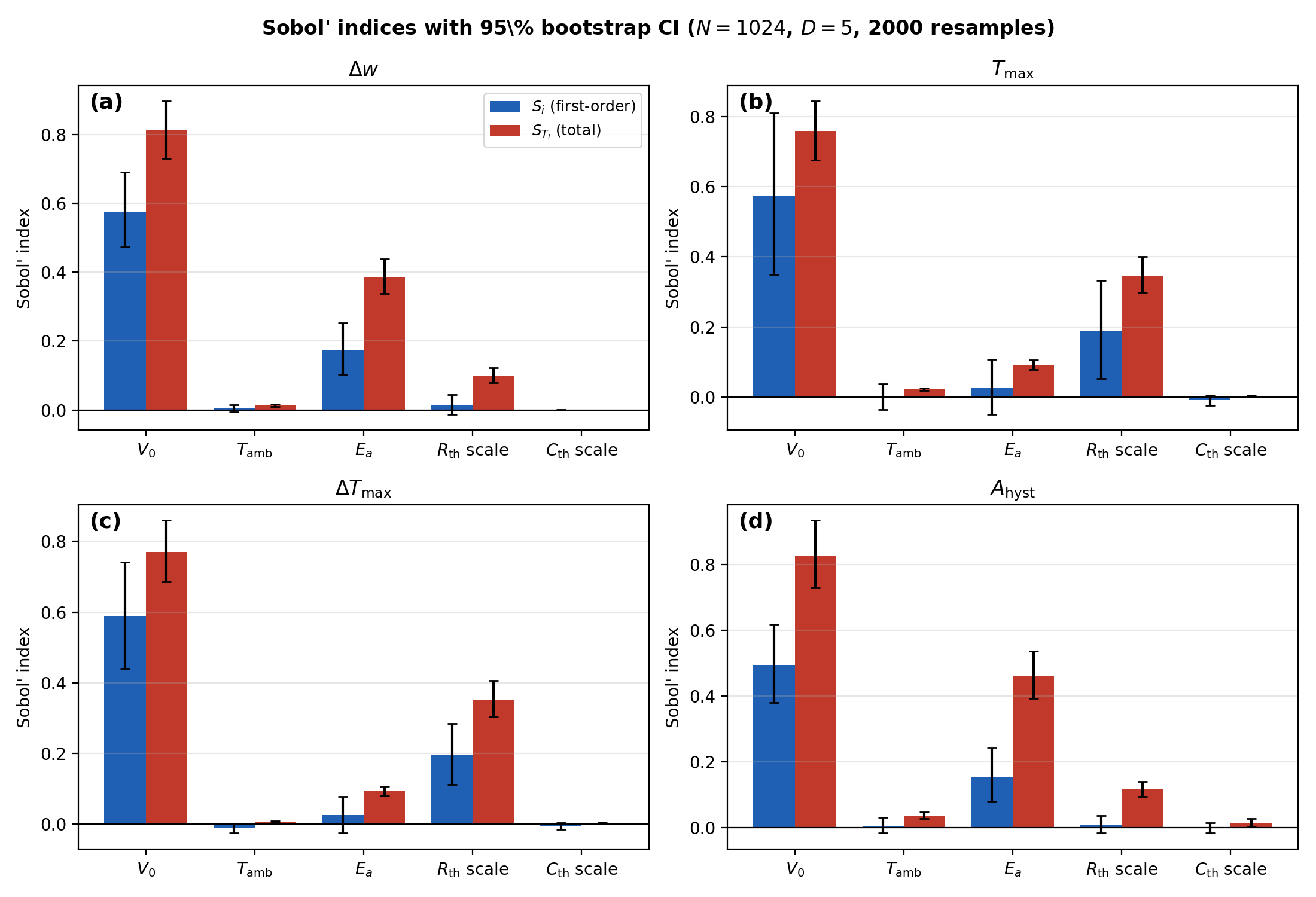}
  \caption{First-order ($S_i$) and total-order ($S_{T_i}$) Sobol'
  indices for the five-parameter ATFM analysis ($N=1024$, $D=5$),
  with $95\%$ bootstrap confidence intervals.}
  \label{fig:sobol-indices}
\end{figure}

Because the Sobol' indices are estimated from a finite Saltelli
ensemble, they are subject to sampling uncertainty. We quantify this uncertainty using a percentile bootstrap with $2000$ resamples of the $N=1024$ base samples, without additional model evaluations. The resulting $95\%$ confidence intervals are shown in \Cref{fig:sobol-indices}. The bootstrap confirms that the small negative first-order estimates obtained for $\Cth$ and, for some outputs, $\Tamb$ are statistically indistinguishable from zero and therefore reflect estimation noise rather than negative physical sensitivities. In contrast, the dominant effects remain well resolved: the total-order indices of $V_0$ range from approximately $0.76$ to $0.83$ across the four outputs, while the contribution of $\Ea$ to
$A_{\mathrm{hyst}}$ remains strongly interaction-dominated.  The excitation amplitude
  $V_0$ dominates all outputs, while $\Ea$ is strongly
  interaction-mediated, $\Rth$ contributes substantially to the thermal response, and $\Cth$ remains negligible.

\begin{figure}[!t]
  \centering
  \includegraphics[width=8.5cm, height=4.5cm,
  trim=0 0 0 0.85cm,clip]{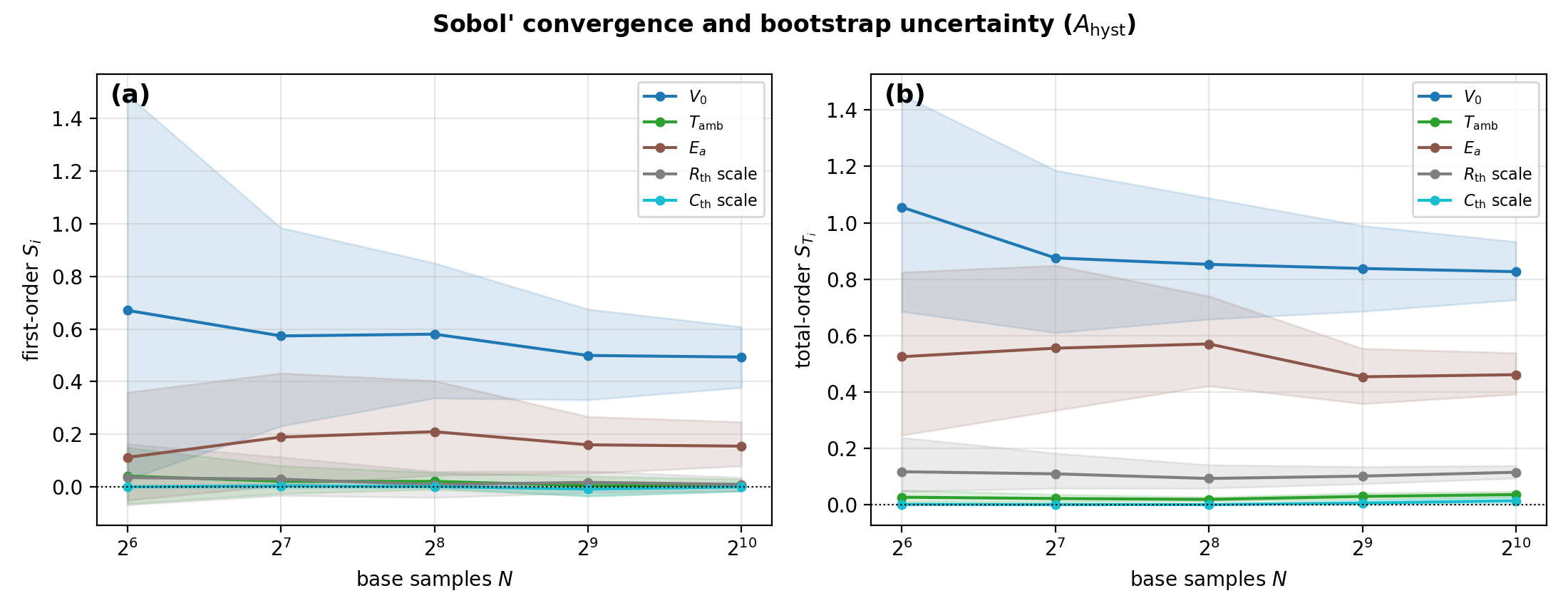}
  \caption{Convergence of the first-order and total-order Sobol'
  indices for $A_{\mathrm{hyst}}$ with the number of base samples
  $N$. Shaded bands denote $95\%$ bootstrap confidence intervals.
  The estimates stabilize with increasing sample size, while the
  associated uncertainty decreases.}
  \label{fig:sobol-convergence}
\end{figure}

The robustness of these estimates is further assessed by repeating the
analysis for
$N\in\{64,128,256,512,1024\}$. As shown in
\Cref{fig:sobol-convergence}, the leading indices progressively
stabilize as the sample size increases, while the associated
confidence intervals decrease. The hysteresis-loop area provides the most demanding case: the total-order estimate of $V_0$ exhibits substantial sampling variability at $N=64$ but approaches a stable value at $N=1024$. More importantly, the parameter ranking remains unchanged from $N=256$ onward, indicating that the reported sensitivity hierarchy is not an artifact of the chosen sample size.

\subsection{Physical Interpretation}
\label{sec:sobol-interpretation}

The global analysis provides a unified picture of the parameter
dependence established by the preceding local and pairwise studies. The excitation amplitude $V_0$ sets the dominant scale of the electrothermal response and controls the location of the
ratchet-to-oscillation transition. The activation energy $\Ea$ becomes important mainly through its coupling to the electrical drive, particularly for the gap excursion and hysteresis-loop area near the transition. The ambient temperature has a comparatively weak global contribution because its strongest effect is localized around the threshold.

Introducing the thermal calibration parameters changes this picture in a specific way. The thermal resistance $\Rth$ emerges as the principal thermal-calibration parameter, directly controlling the temperature rise and influencing the switching dynamics through the Arrhenius kinetics. The thermal capacitance $\Cth$, in contrast, has negligible influence over the investigated range, consistent with the short thermal time-scale and the quasi-static behavior of the thermal state. Thus, the global sensitivity analysis does not simply rank the model parameters. It separates two distinct mechanisms: $V_0$ primarily
controls the strength of the electrothermal excitation, whereas
$\Ea$ and $\Rth$ determine how that excitation is converted into
thermally activated switching. The robustness analysis further shows that this hierarchy is not an artifact of the chosen sample size.
Together, these results identify $V_0$ as the primary control
parameter, $\Ea$ as the key kinetic parameter near the switching
threshold, and $\Rth$ as the principal parameter for thermal
calibration.

\section{Comparison with the Classical Pickett Model}
\label{sec:comparison}
Positioned between compact tunneling-gap and distributed electrothermal filament models, ATFM extends the classical Pickett model with the features summarized in \Cref{tab:comparison}. It preserves the
tunneling-current formulation, series-resistance treatment, and
filament-gap kinetics of the classical Pickett model, while
introducing a dynamic lumped thermal state and an explicit Arrhenius activation factor $\Gamma(T)$. In contrast to distributed electrothermal formulations, the present model does not resolve the spatial temperature field or filament morphology. This deliberate reduction in spatial complexity retains computational efficiency and \texttt{SPICE}-oriented compactness while introducing self-consistent electrothermal feedback into the Pickett switching kinetics.

A further distinguishing feature of ATFM is that the classical
Pickett dynamics are recovered in the isothermal limit
$\Ea\rightarrow0$, providing a direct numerical regression test between the extended and reference formulations. The resulting framework, therefore, separates the contribution of the original tunneling-gap physics from the additional effects introduced by thermal activation, self-heating, and thermal feedback.

\begin{table}[!ht]
\centering
\caption{Feature comparison between the classical Pickett compact model
\cite{Pickett2009,PickettThesis} and ATFM, which directly extends it.}
\label{tab:comparison}
\renewcommand{\arraystretch}{1.2}
\begin{tabular}{@{}p{0.55\linewidth}cc@{}}
\toprule
\textbf{Feature} & \textbf{Pickett} & \textbf{ATFM} \\
\midrule
Physical tunneling gap          & Yes & Yes \\
Quantum-tunneling transport     & Yes & Yes \\
Series-resistance treatment     & Yes & Yes \\
Dynamic temperature             & No  & Yes \\
Lumped thermal RC network       & No  & Yes \\
Arrhenius-activated kinetics    & No  & Yes \\
Electrothermal feedback         & No  & Yes \\
Large-voltage (Kolka) correction& No  & Yes \\
Recovers isothermal limit       & --- & Yes \\
\bottomrule
\end{tabular}
\end{table}
\section{Limitations and Perspectives}
\label{sec:limitations}

The present ATFM formulation provides a computationally efficient
electrothermal description of filamentary resistive switching, but its quantitative applicability is limited by three assumptions.

\begin{enumerate}
\item \textbf{Lumped thermal description.}
The active region is assigned a uniform temperature; local hot spots and spatial heat diffusion are therefore not resolved.

\item \textbf{Effective thermal cross-section.}
Using the junction area as the thermal section is a lumped-model
approximation. Absolute temperatures scale as $\Rth\propto1/A_b$ and should be interpreted as geometry-consistent indicators rather than device-calibrated values. As a calibration parameter, $A_b$ rescales the temperature level, while its physical interpretation as device area reshapes the thermal operating window (\Cref{sec:kolka-window}). The choice $A_b=50\times50$~nm$^2$ is justified in \Cref{tab:geometry}.

\item \textbf{High-temperature range.}
At the largest predicted temperatures, constant material and thermal properties are an approximation. The practical operating window is limited by electrothermal heating and, at larger areas, by kinetic ON-lock (\Cref{sec:kolka-window}).
\end{enumerate}

Future work will focus on experimental calibration, distributed thermal modeling, and validation against measured switching data.

\section{Conclusion}
\label{sec:conclusion}

This work has presented ATFM, an electrothermal extension of the
classical filamentary compact model of Pickett \textit{et al.}
\cite{Pickett2009,PickettThesis} for titanium dioxide resistive switching. The model retains the original tunneling-current formulation, series-resistance regularization, and filamentary switching kinetics while introducing a dynamic thermal balance and an Arrhenius-activated switching-rate factor. In the limit $\Ea\rightarrow0$, ATFM recovers the temperature-independent Pickett dynamics to numerical precision, with direct regression confirming agreement with an independently implemented reference model over excitation amplitude, frequency, and activation energy. The electrothermal model reveals a sharp, current-gated transition between a non-returning ratchet drift and a bounded, drive-locked electrothermal oscillation, arising from the nonlinear exponential locking of the gap kinetics. The excitation amplitude is the dominant control parameter over the investigated domain, whereas the activation energy acts primarily through its interaction with the excitation amplitude near the transition. Peak Joule power and peak temperature exhibit distinct dependencies on excitation amplitude and ambient temperature, highlighting the importance of separating electrical and thermal
control mechanisms.

The correction of the large-voltage ambiguity in the Pickett port
relation provides a monotonic and numerically well-defined tunneling characteristic. The resulting operating-window analysis identifies overheating and, at larger active areas, kinetic ON-lock as the principal constraints on extended operation. Their competition produces a non-monotonic dependence on active area, with an intermediate geometry
maximizing the switching excursion while maintaining a bounded
temperature. This result identifies device geometry as a practical design parameter for balancing switching amplitude and thermal stability. The global sensitivity analysis further establishes a clear hierarchy of model parameters. The excitation amplitude remains the dominant source of output variability, while $\Rth$ is the principal thermal calibration parameter, and $\Ea$ is particularly relevant to the nonlinear switching response near the transition. In contrast, $\Cth$
has negligible influence over the investigated range, supporting the use of the quasi-static thermal reduction when its validity conditions are satisfied. These results provide a quantitative basis for prioritizing voltage control, thermal calibration, and material and geometrical parameters in future device-oriented studies.

ATFM should therefore be regarded as a physically motivated and
computationally tractable electrothermal extension rather than as a fully calibrated device-level model. Experimental calibration of the effective geometry and thermal parameters, together with a distributed electrothermal formulation resolving spatial heat diffusion and filamentary hot spots, remains an important next step. Finally, a fully behavioral \texttt{SPICE} implementation is provided and validated against the
reference model (Appendix~\ref{app:SPICE-validation}), reproducing the $I$--$V$ loop, gap and temperature waveforms, and the full-versus-quasi-static thermal response. Experimental comparison with switching data will provide the next level of device-level validation.

\section*{Appendix}
\begin{appendices}
\section{Complete Governing Equations}\label{appendix}
\renewcommand{\theequation}{\thesection.\arabic{equation}}
\setcounter{equation}{0}
\begin{equation}
\boxed{
\begin{aligned}
\begin{split}
\bullet w_2 = w_1+w-&\tfrac{0.9183}{2.85+4\lambda-2|\Vg|}, \qquad \lambda=L_m/w,\\
\bullet \phi_I = \phi_0-1.15\lambda w&\tfrac{\ln[(w_2/w_1)(w-w_1)/(w-w_2)]}{w_2-w_1}-|\Vg|\tfrac{w_1+w_2}{2w},\\
\bullet I = \sgn(\Vg)&\tfrac{0.0617}{(w_2-w_1)^2}\big[\phi_I e^{-B\sqrt{\phi_I}}-\\
&\qquad \qquad (\phi_I+|\Vg|)e^{-B\sqrt{\phi_I+|\Vg|}}\big], \\
 B=10.246(&w_2-w_1),\\
\bullet V=\Vg+R_sI&, \quad \bullet \Gamma(T)=\exp\!\Big[-\tfrac{\Ea}{\kb}\big(\tfrac1T-\tfrac1{T_0}\big)\Big],\\
\bullet \frac{\dd w}{\dd t}=\Gamma(T)&
\begin{cases}
F_{\mathrm{OFF}}(I,w), & I>0,\\[4pt]
F_{\mathrm{ON}}(I,w), & I<0,
\end{cases}\\
 F_{\mathrm{OFF}}(I,w)&=f_{\mathrm{off}}\,\sinh\!\left(\frac{I}{i_{\mathrm{off}}}\right)\times\\
 \exp\!&\left[-\exp\!\left(\frac{w-a_{\mathrm{off}}}{w_c}-
\frac{|I|}{b}\right)-\frac{w}{w_c}\right],\\
F_{\mathrm{ON}}(I,w)=&-f_{\mathrm{on}}\,\sinh\!\left(\frac{|I|}{i_{\mathrm{on}}}\right)\times\\
 \exp\!&\left[-\exp\!\left(\frac{a_{\mathrm{on}}-w}{w_c}-\frac{|I|}{b}\right)-\frac{w}{w_c}\right].\\
\bullet \Cth\dot T = I\Vg-&\tfrac{T-\Tamb}{\Rth}.  
\end{split}
\end{aligned}}
\label{Eq_Complete}
\end{equation}

\section{\texttt{SPICE} Validation}\label{app:SPICE-validation}
\begin{figure}[!ht]
\centering
\includegraphics[width=9cm,height=10cm,trim=0 0 0 0, clip]{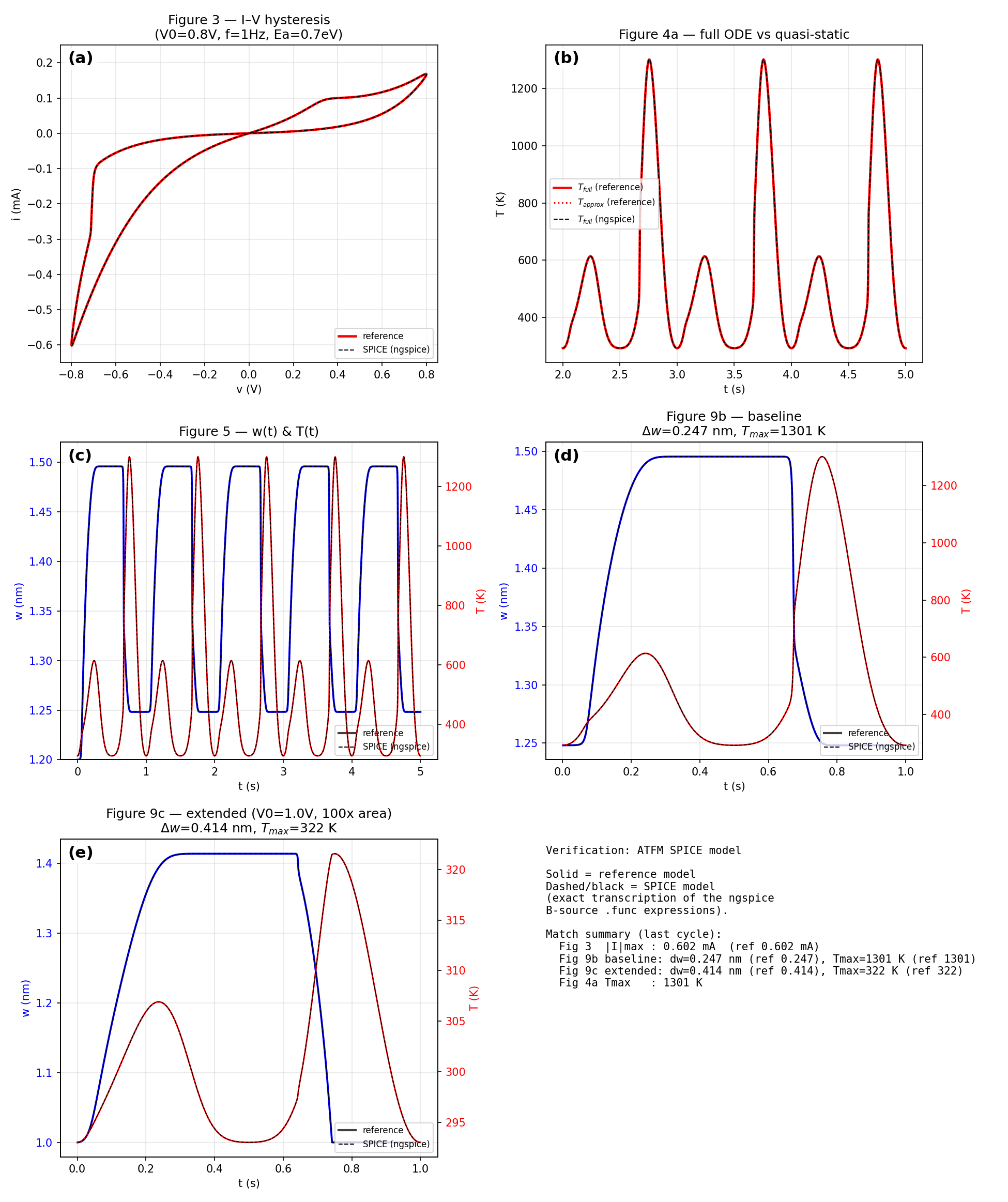}
\caption{\texttt{SPICE} validation of ATFM. Solid curves: reference; dashed black curves: \texttt{SPICE} model (behavioral netlist; Verilog-A/OSDI gives indistinguishable results). (a)~$I$--$V$ hysteresis
($|I|_{\max}=0.602$~mA). (b)~Full-ODE versus quasi-static thermal
response. (c)~Gap $w(t)$ and temperature $T(t)$. (d)~Baseline period ($V_0=0.8$~V, $\Delta w=0.247$~nm, $T_{\max}=1301$~K).
(e)~Extended period ($V_0=1.0$~V, $100\times$ area,
$\Delta w=0.414$~nm, $T_{\max}=322$~K). All \texttt{SPICE} results agree with the reference to the reported precision.}
\label{fig:SPICE-validation}
\end{figure}

To demonstrate the circuit-level usability of ATFM, the complete model was implemented in \texttt{ngspice} using two independent approaches: a dependency-free behavioral netlist and a compiled Verilog-A/OSDI device. Both implementations reproduce the reference Python model at the trajectory level, with the baseline case yielding $|I|_{\max}=0.602$~mA and $T_{\max}=1300.7$~K. Their agreement across all tested operating points is illustrated in \Cref{fig:SPICE-validation}.

\section{Nomenclature}

\Cref{tab:symbols} summarizes the symbols and their units used
throughout the manuscript.

\begin{table}[!ht]
\centering
\caption{List of symbols employed throughout the manuscript.}
\label{tab:symbols}
\begin{tabular}{@{}p{0.27\columnwidth}p{0.75\columnwidth}@{}}
\hline
\textbf{Symbol} & \textbf{Meaning and unit} \\
\hline
$w$,\,\,\,$R_s$ & Tunnel gap [nm],\,\,Series resistance [$\Omega$] \\
$D$,\,$\Vg$ & Oxide thickness [nm],\,Gap voltage [V]  \\
$I$, $V$ & Device current, voltage [A, V] \\
$T$, $\Tamb$, $T_0$ & Device, ambient, and reference temperature [K] \\
$\Rth$, $\Cth$ & Thermal resistance and capacitance [K/W, J/K] \\
$\Ea$ & Activation energy [eV] \\
$\kb$ & Boltzmann constant [J/K (or eV/K)] \\
$\Gamma(T)$ & Thermal activation factor [--] \\
$\phi_0$, $\phi_I$ & Zero-field and mean tunneling barrier [V] \\
$w_1$, $w_2$, $\lambda$, $L_m$ &
Simmons/Pickett auxiliary quantities [nm, nm, --, nm] \\
$a_{\rm on}$, $a_{\rm off}$, $w_c$, $b$ &
Pickett filament-kinetics parameters\\
$f_{\rm on}$, $f_{\rm off}$, $i_{\rm on}$, $i_{\rm off}$ &
Pickett filament-kinetics parameters~[various] \\
\hline
\end{tabular}
\end{table}

\section*{Funding Declaration}
N.G.K. appreciates financial support from the FAPESP--UNESCO-TWAS Project
(Grant No. 2024/08346-8).

\section*{Acknowledgements}
HAC thanks FAPESP grant 2021/14335-0 of the ICTP--SAIFR for partial support.

\section*{Declarations}

\subsection*{Conflict of interest / Competing interests}
The authors declare that they have no conflict of interest and no competing
interests.

\subsection*{Ethics approval and consent to participate}
Not applicable.

\subsection*{Data availability}
Data sharing is not applicable, as no datasets were generated or analyzed. All analyses were performed using Python and  both \texttt{SPICE} (a behavioral \texttt{ngspice} netlist and a \texttt{Verilog-A/OSDI} compact model) implementations.

\section*{Author Contributions}
N.G.K. conceived the study and developed the ATFM formulation, developed the numerical implementation, performed the
electrothermal simulations, carried out the global sensitivity
analysis, and implemented the \texttt{SPICE} and \texttt{Verilog-A/OSDI} models.
N.G.K. performed the data analysis and visualization and drafted the manuscript. All authors contributed to the interpretation of the results, critically reviewed the manuscript, and approved the final version.

\bibliographystyle{snstyle}
\bibliography{Bibliographie}
\end{appendices}
\end{document}